\documentclass[12pt]{iopart2}

\usepackage{epsfig,amssymb,psfig,amsmath}

\def\spose#1{\hbox to 0pt{#1\hss}}
\def\lesssim{\mathrel{\spose{\lower 3pt\hbox{$\mathchar"218$}}
 \raise 2.0pt\hbox{$\mathchar"13C$}}}
\def\gtrsim{\mathrel{\spose{\lower 3pt\hbox{$\mathchar"218$}}
 \raise 2.0pt\hbox{$\mathchar"13E$}}}

\def\<{\langle}
\def\>{\rangle}

\begin{document}

\title{ 
Quantum Ising chains with boundary fields
}

\author{Massimo Campostrini$^1$, 
 Andrea Pelissetto$^2$ and Ettore Vicari$^1$}
\address{$^1$ Dipartimento di Fisica dell'Universit\`a di Pisa
        and INFN, Sezione di Pisa, I-56127 Pisa, Italy}
\address{$^2$ Dipartimento di Fisica di Sapienza, Universit\`a di Roma
        and INFN, Sezione di Roma I, Piazzale A. Moro 2, I-00185 Roma, Italy}

\ead{
campo@df.unipi.it,
Andrea.Pelissetto@roma1.infn.it,
Ettore.Vicari@df.unipi.it}

\begin{abstract}
We present a detailed study of the finite one-dimensional 
quantum Ising chain in a transverse field 
in the presence of boundary magnetic fields coupled with the
order-parameter spin operator.
We consider two  
magnetic fields located at the boundaries of the chain 
that have the same strength and that are  aligned 
in the same or in the opposite direction.
We derive analytic expressions for the gap in all phases for large values 
of the chain length $L$, 
as a function of the boundary field strength. We also investigate 
the behavior of the chain in the quantum ferromagnetic phase 
for oppositely aligned fields, focusing on the
magnet-to-kink transition that occurs at a finite value of the magnetic field 
strength. At this transition we compute 
analytically the finite-size 
crossover functions for the gap, the magnetization profile, 
the two-point correlation function, and the density of fermionic modes. 
As the magnet-to-kink transition is equivalent to the wetting transition in 
two-dimensional classical Ising models, our results provide new 
analytic predictions for the finite-size 
behavior of Ising systems in a strip geometry at this transition.
\end{abstract}

%%%%% \pacs{PACS Numbers: 74.81.Fa, 05.70.Jk, 64.60.Fr}
%74.81.Fa Josephson junction arrays
%64.60.-i general studies of phase transitions
%64.60.Fr Equilibrium properties near critical points
%75.10.Hk classical spin models
%74.78.-w       Superconducting films and low-dimensional structures
%05.10.Ln       Monte Carlo methods
%05.70.Jk       Critical point phenomena 

\maketitle

% ========================= BODY =========================
%\narrowtext

\section{Introduction}
\label{intro}

The quantum Ising chain is a useful theoretical laboratory in which
fundamental issues concerning quantum many-body systems can be thoroughly
investigated, exploiting the exact knowledge of several features of its
phase diagram and quantum correlations. Many results for its
low-energy properties have been derived in the quantum ordered and
disordered phases, and, in particular, at the quantum critical point
separating the two phases, both in the thermodynamic limit and 
in the finite-size scaling (FSS) limit with several types of 
boundary conditions, see, e.g., 
Refs.~\cite{Sachdev-book,NO-11,LSM-61,Katsura-62,%%
Pfeuty-70,BM-71,Suzuki-71,BS-74,CJ-87,BC-87,Henkel-87,BG-87,BT-90,%%
TB-93,IPT-93,Karewski-00,%%
IH-01,Peschel-04,IH-09,DDSCRA-10,CPV-14,CNPV-14,CPV-15} and references therein. 

In this paper we extend these analytic results to the case in which 
boundary fields are present.
We present a detailed study of the quantum Ising chain in 
a transverse magnetic field \cite{Pfeuty-70} in the presence
of magnetic fields coupled with the 
order-parameter spin operator, located at the boundaries of the chain.
We assume the two magnetic fields to have the same strength and consider two
cases: (i) the two fields are parallel; (ii) the two fields are oppositely
aligned. As expected, in the quantum paramagnetic phase the boundary
fields do not change the large-size behavior of low-energies quantities. 
At the critical point, bulk behavior is independent of boundary conditions.
However, the magnetic fields induce a surface phase transition 
with a corresponding scaling behavior. The quantum ferromagnetic phase 
is more interesting. If the boundary fields have opposite direction, one 
observes two different bulk phases. For small magnetic fields, the ground 
state is ferromagnetic as it occurs in the absence of boundary interactions.
On the other hand, kink states \cite{Sachdev-book} are 
the relevant low-energy excitations for large boundary fields.
The two different phases 
are separated by a continuous transition that is only characterized by the 
nature of the two coexisting phases and, indeed, the same transition occurs 
in Ising rings in the presence of a localized link defect \cite{CPV-15}.

Because of the quantum-to-classical mapping, our results can also be applied
to the two-dimensional Ising model,  and, more 
generally, to any model in the Ising universality class, in a strip geometry. 
In the two-dimensional case the quantum transition between the 
ferromagnetic and the kink phase corresponds to the wetting transition, which,
in the context of a strip geometry, is sometimes identified as an 
interface localization/delocalization transition
\cite{Dietrich-88,Indekeu-94,BR-01,BLM-03}.  Our results, therefore,
provide new analytic expressions for two-dimensional Ising systems at the
wetting transition in the presence of boundary magnetic fields.

The paper is organized as follows.  In Sec.~\ref{model} we introduce
the one-dimensional quantum Ising chain with boundary fields. 
In Sec.~\ref{fermmodel}
we compute the low-energy spectrum by exploiting the equivalent
quadratic fermionic formulation of the Hamiltonian 
\cite{LSM-61,Pfeuty-70}.  Explicit
calculations are reported in the following sections.
Secs.~\ref{paramag}, \ref{critpoint} and \ref{magnphase} report
results for the quantum paramagnetic phase, at the quantum critical
point and for the ordered magnetized phase, respectively.
Sec.~\ref{fktra} is devoted to the study of the magnet-to-kink
transition driven by the boundary fields in the ordered phase. 
We obtain exact results for the gap, the magnetization profile, the 
two-point spin-spin correlation function, and the entanglement entropy.
In Sec.~\ref{comp-2D} these results are compared with the existing 
ones for two-dimensional classical models.
Finally,
in Sec.~\ref{conclu} we summarize the main results and draw our
conclusions. A number of appendices report the derivations of some
of the results.

\section{Model and definitions}
\label{model}

The Hamiltonian of the quantum Ising chain in a transverse magnetic
field is given by
\begin{eqnarray}
H  = - J \sum_{i=1}^{L-1} \sigma^{(1)}_i \sigma^{(1)}_{i+1} 
- g \sum_{i=1}^L \sigma^{(3)}_i ,
\label{Isc}
\end{eqnarray}
where $\sigma^{(i)}$ are the Pauli matrices.  In the following we
assume ferromagnetic nearest-neighbor interactions with $J=1$, and
$g>0$. 

The Ising chain undergoes a continuous transition at $g=1$ \cite{Pfeuty-70},
separating a quantum ferromagnetic phase ($g<1$) from a quantum paramagnetic 
phase ($g>1$). In this paper 
we investigate the effects of boundary magnetic fields aligned along
the $x$ axis. They give rise to an additional energy term
\begin{equation}
H_b = - \zeta_1 \sigma_1^{(1)} - \zeta_L \sigma_L^{(1)}
\label{hb}
\end{equation} 
that is added to Hamiltonian (\ref{Isc}).
In the following we shall consider the model in 
the specific case of equal parallel boundary fields (PBF)
\begin{equation}
\zeta_L = \zeta_1 = \zeta,
\label{pbfdef}
 \end{equation}
and of equal oppositely-aligned boundary fields (OBF)
\begin{equation}
\zeta_L = - \zeta_1 = \zeta.
\label{obfdef}
\end{equation}
It is not restrictive to assume $\zeta> 0$ in both cases.

We will often use a basis in which $\sigma_{i}^{(1)}$ is diagonal.
States will be labelled as $|s_1,s_2,\ldots s_L\rangle$, where $s_i$
is the eigenvalues of $\sigma_i^{(1)}$. Signs will be fixed so that
$\sigma_i^{(3)}$ has the form
\begin{equation}
\sigma_i^{(3)} = \begin{pmatrix} 
0 & 1 \\ 1 & 0 \end{pmatrix}
\end{equation}
in this basis.
In the absence of boundary fields, the Hamiltonian commutes with the
generator $P_z = \prod_i \sigma_i^{(3)}$ of the ${\mathbb Z}_2$
transformations $|s\rangle \to -|s\rangle$ and with the operator $T$
of the reflection transformations defined by $T|s_1,s_2,\ldots
s_L\rangle = |s_L,s_{L-1},\ldots, s_1\rangle$. These operators do not
generally commute with $H_b$. However, note that $[P_z T,H_b]=0$ for
OBF.

In this paper we analyze the low-energy spectrum of the model. 
In particular, we compute the
energy differences between the lowest states and the ground state
\begin{equation}
\Delta_{n}\equiv E_n-E_0, \qquad \Delta\equiv \Delta_1,
\label{deltaldef}
\end{equation}
(here 
$E_n$ are the energy eigenvalues ordered so that $E_0\le E_1\le E_2\ldots$, 
and $\Delta$ is the gap), the local magnetization
and the two-point correlation function,
\begin{eqnarray}
m(i) \equiv  \langle \sigma_i^{(1)} \rangle ,\qquad 
G(i,j) \equiv 
\langle \sigma_i^{(1)} \sigma_j^{(1)} \rangle.\label{gxy}
\end{eqnarray}
For OBF, because of the symmetry under $P_z T$, we have 
$m(i) = - m(L-i)$ and, as a consequence, the average magnetization
$\sum_{i=1}^L m(i)$ 
always vanishes.  We also define the integrated correlation $\chi$ and the
correlation length $\xi$ with respect to the center of the chain:
\begin{equation}
\chi = \sum_i G(L/2,i), \qquad
\xi^2 = {1\over 2 \chi} \sum_i (i - L/2)^2 G(L/2,i).\label{chixidef}
\end{equation}

\section{Jordan-Wigner representation and Hamiltonian diagonalization}
\label{fermmodel}

\subsection{Fermionic representation}

To determine the spectrum of Hamiltonian (\ref{Isc}), we extend the
model, considering two additional spins located in 0 and $L+1$ and the
Hamiltonian
\begin{equation}
H_e = - J \sum_{i=1}^{L-1} \sigma_i^{(1)} \sigma^{(1)}_{i+1} - 
    J_0 \,\sigma_0^{(1)} \sigma_1^{(1)} - 
    J_L \,\sigma_{L}^{(1)} \sigma_{L+1}^{(1)}  - 
    g \sum_{i=1}^{L} \sigma_i^{(3)}.
\label{def-Hextended}
\end{equation}
This is the Ising Hamiltonian with two different couplings on the
boundary links and zero transverse field on the boundaries.  Let us
note that the Hamiltonian $H_e$ commutes with both $\sigma_0^{(1)}$ and
$\sigma_{L+1}^{(1)}$, which can therefore be simultaneously
diagonalized.  The Hilbert space can be divided into four sectors, which
we label as $(1,1)$, $(-1,1)$, $(1,-1)$ and $(-1,-1)$, where
$(s_0,s_{L+1})$ are the eigenvalues of $\sigma_0^{(1)}$ and
$\sigma_{L+1}^{(1)}$. The restriction of $H_e$ to each sector gives
rise to the Hamiltonian $H$, defined in Eq.~(\ref{Isc}), with a
boundary term of the form (\ref{hb}).  Hence, the spectrum of $H_e$
also provides the spectrum of $H+H_b$. Let us also note that $H_e$ is
$\mathbb{Z}_2$ symmetric.  Indeed, if $P_z = \prod_{i=0}^{L+1}
\sigma_i^{(3)}$, then $[H_e,P_z] = 0$.  Since $P_z$ does not commute
with $\sigma_0^{(1)}$ and $\sigma_{L+1}^{(1)}$, the spectrum is
necessarily degenerate. Moreover, since $P_z$ anticommutes with these
two boundary operators, $P_z$ maps sector $(1,1)$ to $(-1,-1)$ and
$(1,-1)$ to $(-1,1)$, so that the restriction of $H_e$ to $(1,\pm 1)$
allows us to compute the full spectrum of $H_e$.

To compute the spectrum of Hamiltonian (\ref{def-Hextended}), 
we follow Ref.~\cite{LSM-61}. We first
perform a Jordan-Wigner transformation, defining fermionic operators
$c_i$ and $c_i^\dagger$
\begin{equation}
c_i^\dagger = R_i \sigma_i^+, \qquad c_i = \sigma_i^- R_i ,
\qquad 
R_i = (-1)^{i-1}\prod_{j=1}^{i-1} \sigma_j^{(3)},
\label{c-cdag-def}
\end{equation}
where $\sigma^\pm = (\sigma^{(1)} \pm \sigma^{(2)})/2$. These
relations can be inverted, obtaining
\begin{equation}
\sigma_i^{(1)} = R_i (c_i^\dagger + c_i),  \qquad
\sigma_i^{(3)} = 2 c_i^\dagger c_i - 1 .
\label{sigma-c-relation}
\end{equation}
Thus, Hamiltonian (\ref{def-Hextended}) becomes
\begin{eqnarray}
H_e = -g \sum_{i=1}^L (2 c_i^\dagger c_i - 1) 
    - \sum_{i=0}^L J_i (c_i^\dagger c_{i+1} + c_{i+1}^\dagger c_i + 
     c_i^\dagger c_{i+1}^\dagger + c_{i+1} c_i),
\end{eqnarray}
with $J_i=J=1$ for $i=1,\ldots,L-1$. 
In this formalism $\sigma_0^{(1)} = c_0 + c_0^\dagger$ and 
$\sigma_{L+1}^{(1)} = P_z (c_{L+1}^\dagger - c_{L+1})$. 
To go further, let us rewrite $H_e$ as 
\begin{equation}
H_e = L g - \sum_{i,j=0}^{L+1} \left[c_i^\dagger A_{ij} c_j + 
       {1\over2} c_i^\dagger B_{ij} c_j^\dagger + 
       {1\over2} c_i B_{ij} c_j \right],
\end{equation}
with $A$ and $B$ symmetric and antisymmetric matrices, respectively.
Then, we perform a Bogoliubov transformation. We introduce new
canonical fermionic variables
\begin{equation}
   \eta_k = \sum_{i=0}^{L+1} (g_{ki} c_i^\dagger + h_{ki} c_i),
\end{equation}
where $g_{ki}$ and $h_{ki}$ are fixed by the requirement that $H$ 
takes the form
\begin{equation}
   H_e = E_{gs} + \sum_{k=0}^{L+1} {\cal E}_k \eta^\dagger_k \eta_k,
\label{H1-diag}
\end{equation}
with $0\le {\cal E}_0 \le {\cal E}_1 \le \ldots$
Following Ref.~\cite{LSM-61}, we define the vectors
\begin{eqnarray}
 U_k = (g_{k0} + h_{k0}, g_{k1} + h_{k1},\ldots)  ,\qquad
 V_k = (g_{k0} - h_{k0}, g_{k1} - h_{k1},\ldots)  .
\end{eqnarray}
The variables $\eta_k$ satisfy canonical anticommutation relations if
the vectors $U_k$ form an orthonormal basis, and so does the set
$V_k$.  The vectors $V_k$ satisfy
\begin{equation}
  (A+B)(A-B) V_k = {\cal E}_k^2 V_k,
\end{equation}
which turns the determination of the energies ${\cal E}_k$ into
an eigenvalue problem.  If ${\cal E}_k$ does not vanish, $U_k$ is
given by
\begin{equation}
    U_k = {1\over {\cal E}_k} (A-B) V_k.
\end{equation}
If ${\cal E}_k$ is zero, $U_k$ is the null eigenvector of
$(A-B)(A+B)$. It is also possible to evaluate the constant $E_{gs}$ in
Eq.~(\ref{H1-diag}), which provides the energy of the ground state:
\begin{equation}
   E_{gs} = - {1\over2} \sum_{k=0}^{L+1} {\cal E}_k.
\end{equation}
The matrix $C = {1\over4} (A+B)(A-B)$ is given by (we write 
the $5\times 5$ matrix $C$ for $L=3$,
the generalization to any $L$ being obvious)
\begin{equation}
C = \begin{pmatrix}
    J_0^2   &    g J_0    &     0     &      0      &     0   \\
     g J_0  &  1 + g^2 & g &      0      &     0   \\
      0         &     g      & 1 + g^2 & g   &     0   \\
      0         &     0      & g      & J_L^2 + g^2  & 0    \\
      0         &     0      &     0     &      0 & 0 
     \end{pmatrix} \; .
\label{matrixC-def}
\end{equation}
This matrix $C$ has a zero eigenvalue, ${\cal E}_0 = 0$,
 with eigenvector $V_{0} = (0,\ldots,0,1)$.
Correspondingly, we obtain
$U_{0} = (1,0,\ldots,0)$ and
\begin{eqnarray}
\eta_{0} &=& {1\over2} (c_0 + c_0^\dagger) + 
             {1\over2} (c_{L+1}^\dagger - c_{L+1}) 
= \sigma_0^{(1)} + (-1)^L P_z \sigma_{L+1}^{(1)}
\end{eqnarray}
which depends only on the boundary fermion operators. Note that it can be
rewritten as a combination of $\sigma_0^{(1)}$, $\sigma_{L+1}^{(1)}$,
and $P_z$, hence it does not represent an additional symmetry of the
Hamiltonian. The presence of a zero eigenvalue was expected, as the
spectrum is degenerate.  If $J_0$ and $J_L$ do not vanish, since
there is only one fermion operator with zero energy, the spectrum is
doubly degenerate.

Let us now consider the nonzero eigenvalues. Since the vectors $V_k$
are an orthonormal set, orthogonality with $V_{0}$ implies $V_k =
(a_0,\ldots,a_L,0)$. It follows analogously $U_{k} =
(0,b_1,\ldots,b_{L+1})$. Therefore, we have for $k=1,\ldots,L+1$
\begin{equation}
\eta_k = {a_0\over2} (c_0^\dagger - c_0) + 
         {b_{L+1}\over 2} (c_{L+1} + c_{L+1}^\dagger) + \ldots
\end{equation}
where the dots represent a polynomial in $c_j$ and $c_j^\dagger$ with 
$1\le j \le L$. It is then easy to verify that 
\begin{equation}
\{\eta_k,\sigma_0^{(1)}\} = [\eta_k,\sigma_{L+1}^{(1)} ] = 0.
\end{equation}
It follows that, if a state $|\psi\rangle$ satisfies
$\sigma_0^{(1)}|\psi\rangle = s_0 |\psi\rangle$, then we have
$\sigma_0^{(1)}\eta_k |\psi\rangle = -s_0 \eta_k |\psi\rangle$.
Analogously, if $\sigma_{L+1}^{(1)}|\psi\rangle = s_1 |\psi\rangle$,
we have $\sigma_{L+1}^{(1)}\eta_k |\psi\rangle = s_1 \eta_k
|\psi\rangle$.  Therefore, if $|\psi\rangle$ belongs to sector
$(s_0,s_{L+1})$, then $\eta_k |\psi\rangle$ belongs to sector
$(-s_0,s_{L+1})$.

To conclude, we should determine to which sectors the two degenerate
ground states belong. If $J_0$ and $J_L$ are both positive, we expect
the ground-state configurations to be ordered. Therefore, if on one
boundary the spin is directed in the positive $x$ direction, we expect
the same to occur at the other end of the chain. Therefore, we
conclude that the ground states belong to sectors $(+1,+1)$ and
$(-1,-1)$.  This identification is supported by exact diagonalization.

We can now classify the states. We only consider the states with $s_0
= +1$, to avoid the double degeneracy. Then, if $J_0, J_L > 0$ we
have:
\begin{itemize}
\item[1)] In the sector $s_0 = 1$, $s_{L+1} = 1$, the lowest energy
  state is the ground state of the Hamiltonian and all states are
  obtained as $\eta_{k_1}^\dagger \ldots \eta_{k_m}^\dagger
  |0\rangle$, $k_i \ge 1$, with $m$ even.  The first excited state is
  $\eta_2^\dagger\eta_1^\dagger|0\rangle$ and the energy gap is
  $\Delta = {\cal E}_1 + {\cal E}_2$.
\item[2)] In the sector $s_0 = 1$, $s_{L+1} = -1$, the lowest energy
  state is the first excited state $\eta_1^\dagger|0\rangle$ of the
  Hamiltonian $H_e$ and all states are obtained as $\eta_{k_1}^\dagger \ldots
  \eta_{k_m}^\dagger |0\rangle$, $k_i \ge 1$, with $m$ odd. 
In particular, the first
  excited state in the sector is $\eta_2^\dagger|0\rangle$, so that
  the energy gap is $\Delta = {\cal E}_2 - {\cal E}_1$.
\end{itemize}
The first case is relevant when considering $\zeta_1,\zeta_L > 0$ in
the boundary Hamiltonian $H_b$, which are identified with $J_0$ and
$J_L$, respectively. The second case is relevant when $\zeta_1$ and
$\zeta_L$ have opposite signs: the corresponding gap is obtained by
identifying $\zeta_1=J_0$ and $\zeta_L=- J_L$.

\subsection{Exact results for the energy gap}

The energy spectrum of Hamiltonian $H_e$ can be computed exactly, as
discussed in \ref{App.A}. Here we will focus on the case
$J_0=J_L=\zeta$, which allows us to derive the spectrum for parallel
or opposite equal magnetic fields at the boundary, cf.
Eqs.~(\ref{pbfdef}) and (\ref{obfdef}), respectively.  Neglecting the
irrelevant zero eigenvalue, we obtain $L+1$ elementary excitations
whose energy ${\cal E}_m$ is given by
\begin{equation}
{\cal E}_m = 2 \sqrt{1 + g^2 - 2 g \cos k_m},
\end{equation}
where $k_m$ are the $L+1$ solutions of the equation
\begin{eqnarray}
&&[(1 + g^2)(1 - 2 \zeta^2) + \zeta^4 - 2 g \cos k (1 - \zeta^2)^2] \sin k L
= \label{secular-equation}\\
&&\quad g (1 - 2 g \cos k + g^2 - \zeta^4) \sin [k (L+1)].
\nonumber
\end{eqnarray}
We should consider real solutions in $[0,\pi[$ (correspondingly $|1 -
    g| \le {\cal E}_m \le 1 + g$), purely imaginary solutions $k = i
    h$ with $h > 0$ (correspondingly ${\cal E}_m < |1-g|$), and
    solutions $k = \pi + i h$ (correspondingly ${\cal E}_m > 1+g$).
    Equation~(\ref{secular-equation}) has also a spurious solution for
    $k=0$ for all values of the parameters, which should be
    discarded. For $g = 1-\zeta^2$, there is also a second solution
    with $k = 0$, which corresponds to a true excitation of the model.
    For $\zeta = 0$ we obtain the equation appropriate for open
    boundary conditions (OBC) \cite{Pfeuty-70},
\begin{equation}
{\sin k (L+1)\over \sin k L} = {1\over g},
\end{equation}
and $1 + g^2 - 2 g \cos k = 0$, which implies an additional zero mode
for $H_e$. This is not surprising as the spectrum has a fourfold
degeneracy for $J_0 = J_L = 0$: the four sectors are
equivalent in the absence of boundary fields.

The real solutions of Eq.~(\ref{secular-equation}) are obtained by
solving
\begin{equation}
\tan k L = {g \sin k (1 + g^2 - \zeta^4 - 2 g \cos k) \over 
  (1 - \zeta^2)^2 + g^2 (1 - 2 \zeta^2) - 
  g [g^2 - 1 + (2 - \zeta^2)^2]\cos k + 2 g^2 \cos^2 k}.
\label{realtankL}
\end{equation}
The imaginary solutions are obtained by setting $k = i h$.
Eq.~(\ref{secular-equation}) then becomes
\begin{equation}
e^{2 h L} = {e^{-h} (e^h g - 1) [e^h (1 - \zeta^2) - g]^2 \over
             (g - e^h) (e^h g - 1 + \zeta^2)^2 }.
\label{equazione-J1eqJ2-small}
\end{equation}
The corresponding energies ${\cal E} = 2 \sqrt{1 + g^2 - 2 g \cosh h}$
are always smaller than those corresponding to real momenta, hence the
localized states are the most important ones in the determination of
the low-energy spectrum.

\section{The quantum paramagnetic case} 
\label{paramag}

For $g > 1$ (paramagnetic phase) the lowest energy state is a
localized state, i.e., a solution of
Eq.~(\ref{equazione-J1eqJ2-small}) with $h > 0$.  For large values of
$L$ we must consider the poles in the right-hand side of the equation.
They are
\begin{eqnarray}
e^h =g,\qquad
e^h = {1 - \zeta^2\over g}.
\label{equazione-poli}
\end{eqnarray}
For $g > 1$, the second equation does not have any solution. The
relevant pole is the first one, so that the relevant solution can be
written as $e^h = g + \epsilon$. For large values of $L$, 
substituting this expression in Eq.~(\ref{equazione-J1eqJ2-small}), we obtain
\begin{equation}
\epsilon = - {\zeta^4 g (g^2 - 1)\over (g^2 - 1 + \zeta^2)^2} g^{-2 L}.
\end{equation}
Correspondingly, we find
\begin{equation}
{\cal E}_1 = 2 {\zeta^2 (g^2 - 1) \over (g^2 - 1 + \zeta^2)} g^{-L}.
\label{calE1_para}
\end{equation}
The energy ${\cal E}_1$ vanishes with exponentially small corrections.
The energy of the second excited state corresponds to the lowest value
$k_{\rm min}$ of $k$ satisfying Eq.~(\ref{realtankL}). We obtain
$k_{\rm min} = \pi/L + O(L^{-2})$ and
\begin{equation}
{\cal E}_2 = 2 (g-1) + {g \pi^2 \over g - 1} {1\over L^2} + O(L^{-3}).
\end{equation}
The system is therefore gapped, and the same gap is obtained for both
PBF and OBF modulo exponentially small corrections, since ${\cal E}_2
- {\cal E}_1 \approx {\cal E}_2 + {\cal E}_1 + O(g^{-L})$:
\begin{equation}
\Delta = 2 (g-1) + {g \pi^2 \over g - 1} {1\over L^2} + O(L^{-3}).
\end{equation}
Note that the same result would have been obtained for OBC,
confirming that boundary conditions and/or boundary fields
are irrelevant in the paramagnetic
quantum phase.

\section{Critical-point behavior}
\label{critpoint}

\subsection{Finite-size scaling for $ g = 1$}

At the critical point $g = 1$, there are no localized solutions, hence
the two lowest energies ${\cal E}_n$ are obtained by considering
Eq.~(\ref{realtankL}). The equation becomes
\begin{equation}
\tan k L = {\cot(k/2) (\zeta^4 - 4 \sin^2 k/2) \over 
            4 \zeta^2 - \zeta^4 - 4 \sin^2 {k/2} }.
\label{seceq-g1}
\end{equation}
The lowest-energy solutions correspond to momenta that scale as $k
\sim 1/L$. Therefore, we can expand the right-hand side in powers of
$k$, obtaining
\begin{equation}
\tan k L = {2 \zeta^2\over 4 - \zeta^2} {1\over k} + O(k),
\end{equation}
which shows that $\tan k L$ must diverge for $L\to \infty$. This fixes
$k_n \approx (n - 1/2) \pi/L$ with $n=1,2,\ldots$ Including the
corrections, we obtain for the two lowest momenta
\begin{eqnarray}
  k_1 ={\pi\over 2 L} + {\pi\over L^2} {\zeta^2 - 4\over 4 \zeta^2}, \qquad 
  k_2 = {3 \pi\over 2 L} + {3 \pi\over L^2} {\zeta^2 - 4\over 4 \zeta^2}, 
\end{eqnarray}
so that 
\begin{eqnarray}
{\cal E}_1 = {\pi\over L} + {\pi\over L^2} {\zeta^2 - 4\over 2 \zeta^2}, \qquad
{\cal E}_2 = {3\pi\over L} + {3 \pi\over L^2} {\zeta^2 - 4\over 2 \zeta^2}. 
\end{eqnarray}
These results imply that the PBF and OBF gaps are 
\begin{eqnarray}
&&\Delta_{\rm PBF} = {4\pi\over L} + 
{2 \pi\over L^2} {\zeta^2 - 4\over \zeta^2},
\label{depbfc}\\
&&\Delta_{\rm OBF} = {2\pi\over L} + {\pi\over L^2} {\zeta^2 - 4\over \zeta^2},
\label{depofc}
\end{eqnarray}
respectively.
Note that the amplitude of the $1/L$ term 
differs from that obtained in the OBC case (corresponding to
$\zeta=0$) \cite{Pfeuty-70}, 
\begin{equation}
\Delta_{\rm OBC} = {\pi\over L} +O(L^{-2}), 
\label{deltaobccr}
\end{equation}
indicating that $\zeta = 0$ is a surface critical point.

As we have discussed in Ref.~\cite{CPV-14} in the OBC case, 
the corrections of order
$1/L$ to the leading behavior can be interpreted as the effect of a
nonlinear scaling field associated with $L$: the leading irrelevant 
boundary operator gives corrections that scale as $L^{-2}$. 
The same holds in the presence of boundary fields.
Indeed, we can define a
new length scale $L_{\rm eff}$ so that
\begin{equation}
\Delta_{\rm PBF} = {4\pi\over L_{\rm eff}} + O(L^{-3}_{\rm eff}), \qquad
\Delta_{\rm OBF} = {2\pi\over L_{\rm eff}} + O(L^{-3}_{\rm eff}).
\end{equation}
Using the previously reported expressions we obtain 
\begin{equation}
L_{\rm eff} = L - {1\over2} {\zeta^2 - 4\over \zeta^2}.
\label{Leff-def}
\end{equation}
Note that the same rescaling applies both to PBF and OBF, a result
which is not obvious in the original formulation, but which has
a natural explanation in the present setting in which we consider $H_e$.
Indeed, the rescaling should cancel the leading correction in the
whole low-energy spectrum of $H_e$, hence in all low-energy excitation energies
${\cal E}_k$. Therefore, it should apply to both PBF and OBF.

\begin{figure}[tbp]
\centerline{\epsfig{width=10truecm,angle=-90,file=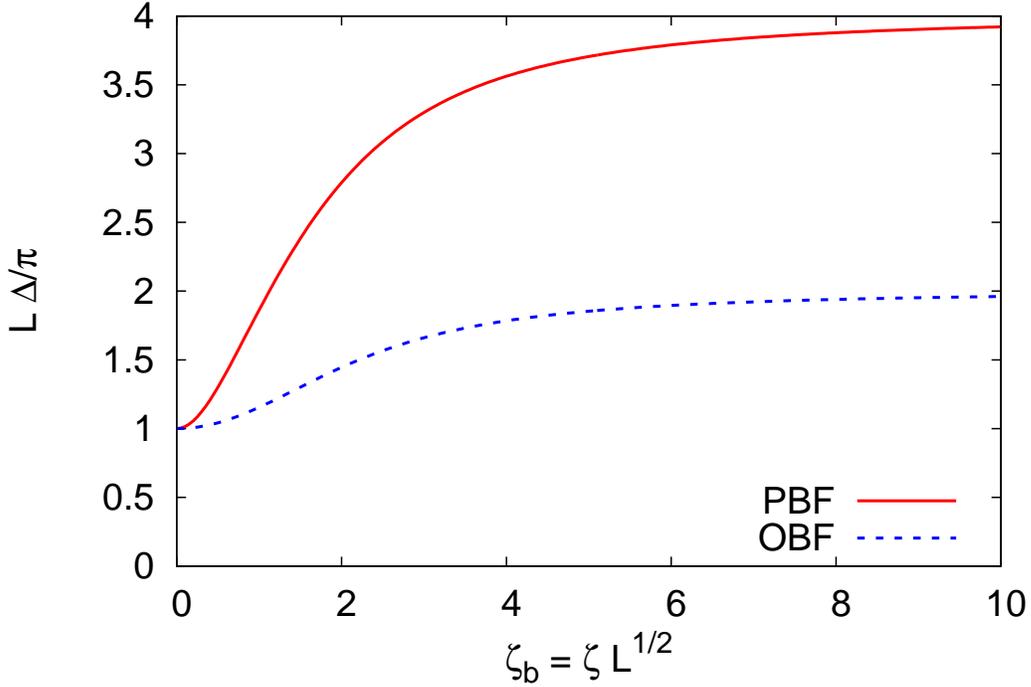}} 
%=% \vskip-5mm
\caption{(Color online) Crossover curves for $L \Delta/\pi$ as a function 
  of $\zeta_b = \zeta L^{1/2}$. For $\zeta_b \to 0$ we recover the OBC 
  case and $L \Delta/\pi \to 1$. For $\zeta_b \to +\infty$ we obtain 
  $L \Delta/\pi \to 2$  for OBF and $L \Delta/\pi \to 4$ for PBF.
}
\label{FSS-crossover-etab}
\end{figure}

Since $\zeta = 0$ is a surface critical point and the boundary term is 
a relevant perturbation, we can study the crossover behavior close to 
$\zeta = 0$. The relevant scaling variable is $\zeta_b = \zeta L^{1/2}$.
Therefore, in the limit $\zeta\to 0$, $L\to \infty$ at fixed $\zeta_b$ 
we have 
\begin{equation}
   \Delta(\zeta,L) = L f_b(\zeta_b),
\end{equation}
where the function $f_b(\zeta_b)$ depends on the type, OBF or PBF, of 
boundary fields. In the scaling limit defined above, momenta 
scale as $k = b/L$. Using Eq.~(\ref{seceq-g1}), we obtain 
an implicit equation for $b$:
\begin{equation}
   \tan b = {1\over 2 b} {\zeta_b^4 - b^2 \over \zeta_b^2}.
\label{scaling-b-boundary}
\end{equation}
Therefore, to compute the scaling function, 
one determines the two lowest values of $b$ satisfying
Eq.~(\ref{scaling-b-boundary})
and uses them to determine the gap as before. The scaling 
curves are reported in Fig.~\ref{FSS-crossover-etab}.

\subsection{Scaling behavior close to the critical point}

Let us now consider the scaling behavior in a neighborhood of the
critical point $g = 1$. The corresponding scaling variable is
\begin{equation}
 w = (g - 1) L.
\end{equation}
The relevant equations are obtained from Eqs.~(\ref{realtankL})
and (\ref{equazione-J1eqJ2-small}) by expanding the right-hand sides 
in the limit $g\to 1$, $L\to \infty$, $k\to 0$ at fixed $w$ and $k L$.
As expected, as long as $\zeta\not=0$, we obtain a result that is independent
of $\zeta$: the FSS behavior only depends on the nature of the boundary 
conditions, but not on the specific values of the boundary fields.

A localized state exists only for $w \ge 1$. If we set $h = \delta/L$,
the parameter $\delta$ is a solution of the equation
\begin{equation}
\delta = w \tanh \delta.
\end{equation}
The corresponding excitation energy is 
\begin{equation}
{\cal E}_{\rm loc} = {2\over L} \sqrt{w^2 - \delta^2}.
\end{equation}
  Note that $\delta \to 0$ for $w \to 1$, so
that ${\cal E}_{\rm loc} L \to 2$ in the limit.  On the other hand,
for $w\to \infty$ we obtain $\delta\to w (1 - 2 e^{-2w}) $ so that
${\cal E}_{\rm loc} L \approx 4 w e^{-w}$, consistently with
Eq.~(\ref{calE1_para}).

\begin{figure}[tbp]
\centerline{\epsfig{width=10truecm,angle=-90,file=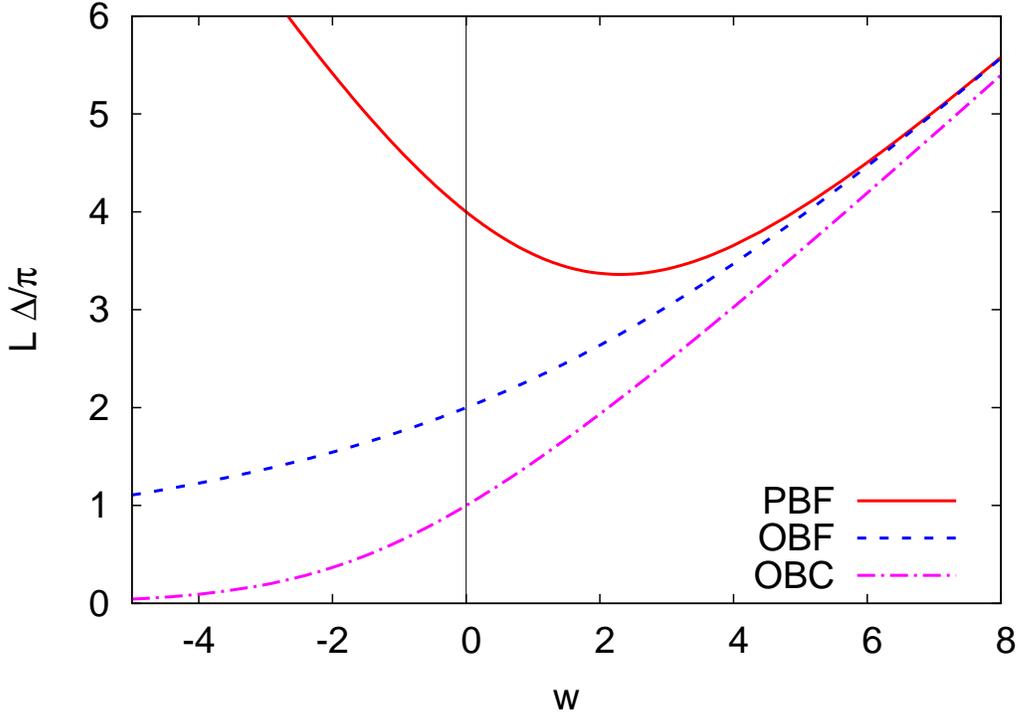}}
%=% \vskip-5mm
\caption{(Color online) FSS curves for 
  $L \Delta/\pi$, for PBF, OBF, and OBC, as a function of $w$.
  For $w = 0$, $L \Delta/\pi= 4,2,1$ in the three cases, respectively.
  }
\label{FSS-Delta1}
\end{figure}

The propagating modes can be written as $k = \epsilon_n/L$, where
$\epsilon_n$ are the positive solutions of the equation
\begin{equation}
\epsilon_n = w \tan \epsilon_n,
\label{equation_epsilon}
\end{equation}
with $(n-3/2) \pi \le \epsilon_n \le (n - 1/2) \pi$ (the solution with $n=1$
only exists for $0 <  w \le 1$). The corresponding excitation energies are 
then ${\cal E}_n = 2 \sqrt{w^2 + \epsilon_n^2}$. 
Using these results we obtain for the energy gap in the FSS limit:
\begin{eqnarray}
L \Delta(w) = 2 \sqrt{w^2 + \epsilon_3^2} \pm 2 \sqrt{w^2 + \epsilon_2^2} &&
     \qquad \hbox{for } w \le 0, \nonumber \\
L \Delta(w) = 2 \sqrt{w^2 + \epsilon_2^2} \pm 2 \sqrt{w^2 + \epsilon_1^2} &&
     \qquad \hbox{for } 0 < w \le 1, \nonumber \\
L \Delta(w) = 2 \sqrt{w^2 + \epsilon_2^2} \pm 2 \sqrt{w^2 - \delta^2} &&
     \qquad \hbox{for } w > 1.
\label{gap-gcritico}
\end{eqnarray}
The two signs refer to the case of parallel (plus sign) and
opposite (minus sign) boundary fields.  The curves are reported in
Fig.~\ref{FSS-Delta1} together with the corresponding OBC scaling
function.  In the paramagnetic phase all curve have a similar shape
--- they coincide asymptotically --- while in the phase $g < 1$ ($w < 0$), the
behavior is clearly different. For $w \to -\infty$, we 
have $L \Delta_{\rm PBF} (w) \sim 4 |w|$, 
$L \Delta_{\rm OBF} (w) \sim 3 \pi^2/|w|$, and 
$L \Delta_{\rm OBC} (w) \sim 4 |w| e^w$.

It is interesting to discuss the scaling corrections, that can be
easily shown to decay as $1/L$, as it occurs for $\zeta = 0$. In
Ref.~\cite{CPV-14} we argued that such corrections are not associated
with irrelevant boundary operators, which give rise to corrections of order
$L^{-2}$, but that they should be interpreted as due to nonlinear scaling
fields. The same holds for $\zeta\not=0$. 
First, one should consider the effective length $L_{\rm eff}$
defined in Eq.~(\ref{Leff-def}). Then, one should consider the
nonlinear scaling field $u_\mu$ associated with $g-1$ and, finally,
rescale the energies with the {\em sound velocity} $c(g)$.  The
nonlinear scaling fields do not depend on the boundary conditions,
hence we can use the expressions reported in Ref.~\cite{CPV-14}:
\begin{equation}
c(g) = 2 \sqrt{g} \qquad u_\mu = (g-1)/\sqrt{g}.
\end{equation}
Then, we define the new scaling variable
\begin{equation}
\widetilde{w} = u_\mu L_{\rm eff}
\end{equation}
and express $\epsilon_n$ and $\delta$ as functions of
$\widetilde{w}$. A somewhat lengthy calculation shows that $c(g)
L \Delta/2$ can be written as in Eq.~(\ref{gap-gcritico}), with
$\widetilde{w}$ replacing $w$, with corrections that scale as $L_{\rm
  eff}^{-2}$. As expected, the leading corrections cancel out.

\section{The magnetized phase} \label{sec5}
\label{magnphase}

Let us now consider the case $g < 1$. In the infinite-volume limit,
there exist two degenerate ground states $| + \rangle$ and $|
-\rangle$, which differ by the value of the
magnetization~\cite{Pfeuty-70}
\begin{eqnarray}
m_\pm = {\rm lim}_{h\to 0^\pm} {\rm lim}_{L\to\infty} 
\langle \sigma_i^{(1)} \rangle = \pm m_0, \qquad
m_0=(1-g^2)^{1/8},
\label{modef}
\end{eqnarray}
where $h$ is a bulk magnetic field applied along the $x$ direction.  For a
chain of finite size $L$, the degeneracy is lifted. In the OBC case,
the energy difference between the two lowest-energy states vanishes
exponentially as $L$ increases \cite{Pfeuty-70}:
\begin{eqnarray}
&&\Delta_{\rm OBC} \equiv E_1-E_0 = 2 (1-g^2) g^L\left[1+ O(g^{2L})\right]. 
\label{deltaobc}
\end{eqnarray}
Moreover, $m(x)=0$ by symmetry, and 
\begin{equation}
{G(L/2,x)\over m_0^2}\to 1\quad{\rm for}\;L\to\infty.
\label{gm0l}
\end{equation}
These are the standard features of a quantum ferromagnetic phase with
a spontaneously broken $\mathbb{Z}_2$ symmetry.

We now extend the discussion to the case in which boundary fields
are present, considering equal PBF and OBF. To
identify localized states, we consider
Eq.~(\ref{equazione-J1eqJ2-small}). The relevant solutions are
associated with the poles in the right-hand side of the equation, see
Eq.~(\ref{equazione-poli}). It is immediate to verify that there are
no poles for $\zeta^2 > 1 - g$, while a pole exists in the opposite
case. Therefore, we distinguish three different cases, depending on
whether $\zeta$ is smaller, larger, or equal to 
\begin{equation}
\zeta_c(g) = (1 - g)^{1/2}.
\label{zetacg}
\end{equation}

\subsection{Spectrum}

\subsubsection{Low-field regime, $\zeta < \zeta_c(g)$.}

For $\zeta < \zeta_c(g)$, the two lowest-energy excitations corresponding
to  ${\cal E}_1$
and ${\cal E}_2$ are localized.  We have $k = ih$ ($h > 0$) with
\begin{equation}
e^h = s (1 + \delta), \qquad s = {1 - \zeta^2\over g},
\end{equation}
where $s > 1$ and $\delta$ is a correction term. Let us first obtain a
large-$L$ expression for $\delta$ that is uniform as $\zeta \to 0$. 
This requires particular
care as the right-hand side of Eq.~(\ref{equazione-J1eqJ2-small})
has a different singular behavior for $\zeta = 0$ and $\zeta
\not=0$. For $\zeta = 0$, it has a simple pole as $e^h \to s = 1/g$. On
the other hand, for $\zeta \not=0$, it has a double pole for $e^h \to s$.

To compute $\delta$ we set $e^h = s$ in all terms of
Eq.~(\ref{equazione-J1eqJ2-small}) that are not relevant for the
singular behavior as $L\to \infty$, i.e., we write
\begin{equation}
s^{2 L} = {(1 - g e^h) \over C_1 g (e^h - s)^2}.
\label{eq-magn-1}
\end{equation}
with
\begin{equation}
C_1 = {s(s-g)\over g (s^2 - 1)^2}.
\end{equation}
Equation (\ref{eq-magn-1}) is a second-order algebraic equation of $e^h$,
whose solutions are
\begin{equation}
e^h = s + {s^{-2L}\over 2 C_1 g} \left(\pm \sqrt{4 C_1 g \zeta^2 s^{2L} + g^2} - g\right).
\end{equation}
The energies ${\cal E}^2/4 = 1 + g^2 - 2 g \cosh h$ are correspondingly
\begin{equation}
{1\over 4} {\cal E}_\pm^2 = {\zeta^2 (1 - g^2 - \zeta^2)\over (1 - \zeta^2)}
   - {g (1 - s^2)^3 s^{-2L} \over 2 (g-s) s^3}
    \left(\pm \sqrt{4 C_1 g \zeta^2 s^{2L} + g^2} - g\right)
\label{Epm-general}
\end{equation}
and ${\cal E}_1 = {\cal E}_+$, ${\cal E}_2 = {\cal E}_-$.

There are now two interesting cases. First, we take the limit
$L\to\infty$ at fixed $\zeta \not = 0$. We can then simplify 
Eq.~(\ref{Epm-general}), obtaining
\begin{equation}
{1 \over 4} {\cal E}_\pm^2 = {\zeta^2 (1 - g^2 - \zeta^2)\over (1 - \zeta^2)}
   \pm \zeta {g (1 - s^2)^3 s^{-L} \over (s-g) s^3}
    \sqrt{C_1 g}.
\end{equation}
These results imply
\begin{eqnarray}
\Delta_{\rm PBF} \equiv {\cal E}_2 + {\cal E}_1 &= &
4 \zeta \sqrt{1 - g^2 - \zeta^2\over 1 - \zeta^2} + O(s^{-L}),
\\
\Delta_{\rm OBF} \equiv {\cal E}_2 - {\cal E}_1 & = &
{ 2 g (s^2-1)^{2} \over (s-g) s^2} \; s^{-L} + O(s^{-2L}).\quad
\label{deltalf}
\end{eqnarray}
In the PBF case the gap is finite and proportional to $\zeta$ 
for small fields. Instead, $\Delta_{\rm OBF}$
vanishes exponentially, as in the OBC case (which corresponds to
vanishing boundary fields).

A second possibility consists in considering the FSS limit around
$\zeta = 0$. As discussed in Ref.~\cite{CNPV-14}, the ratio
$\Delta_{\rm PBF}(\zeta)/\Delta_{\rm OBC}$ is expected to become a
function of $\kappa = \zeta/\Delta_{\rm OBC}$ (to compare with the
formulae of Ref.~\cite{CNPV-14} note that the energy associated with
the perturbation is here proportional to $\zeta$, without additional
factors of $L$). In this limit $s^L \zeta \approx g^L \zeta$ is 
proportional to $\kappa$. 
Using the expression for $\Delta_{\rm OBC}$ reported 
in Eq.~(\ref{deltaobc}), we obtain
\begin{equation}
{{\cal E}_\pm \over \Delta_{\rm OBC}}  = 
   {1\over 2} \sqrt{1 + 16 (1 - g^2) \kappa^2} \mp {1\over2},
\end{equation}
so that 
\begin{equation}
{\Delta_{\rm PBF}(\zeta) \over \Delta_{\rm OBC}} = 
   \sqrt{1 + 16 (1 - g^2) \kappa^2}.
\end{equation}
Note that the scaling function agrees with that predicted in
Ref.~\cite{CNPV-14}. We can also consider the ratio $\Delta_{\rm
  OBF}(\zeta)/\Delta_{\rm OBC}$ in the same limit, but in this case we
obtain that the ratio is simply equal to 1, i.e., independent of
$\kappa$.  The reason can be easily understood. For OBF the energy
associated with the perturbation is zero---the contributions of the
two boundaries cancel---hence $\kappa = 0$ identically.

\subsubsection{Large-field phase, $\zeta > \zeta_c(g)$.}

For $\zeta > \zeta_c(g)$ there are no localized states.  
The secular equation 
(\ref{realtankL}) can be rewritten as
\begin{equation}
\tan kL = \sin k f(k,\zeta,g),\qquad
f(0,\zeta,g) \not = 0.
\label{taneq}
\end{equation}
For $L\to \infty$, the solutions can be parametrized as
\begin{equation}
k_n = {\pi n\over L} + {a_n\over L^2}.
\end{equation}
It follows that
\begin{equation}
\tan {a_n \over L} \approx  \sin \left({\pi n\over L}  + {a_n\over L^2} \right)
       f(k_n,\zeta,g).
\end{equation}
For large values of $L$ we obtain finally
\begin{equation}
a_n = a n =
  {\pi n} f(0,\zeta,g) = \pi n {g (1 - g + \zeta^2)\over (1-g)(1-g-\zeta^2)}.
\end{equation}
Note that the correction term $a_n$ diverges as $\zeta \to \zeta_c(g)$
and has a finite limit as $\zeta \to \infty$. Moreover, the explicit
form of the Hamiltonian is only relevant for the corrections. It
follows
\begin{equation}
{\cal E}_n = {2(1-g)} + {g\over 1 - g} {\pi^2 n^2\over L^2} +
             {2 a g\over 1 - g} {\pi n^2\over L^3} + O(L^{-4}).
\end{equation}
We obtain finally 
\begin{eqnarray}
\Delta_{\rm PBF} & = & 4 (1 - g) + {5 g\over 1-g} \, {\pi^2\over L^2}
+ O(L^{-3}) ,
\\
\Delta_{\rm OBF} & = & {3 g\over 1-g} \, {\pi^2\over L^2}
+ {6 g^2 (1-g + \zeta^2) \pi^2 \over (1-g)^2 (1-g-\zeta^2) L^3 } + O(L^{-4}).
\label{deltalk}
\end{eqnarray}
For OBF, the gap vanishes for  $L\to \infty$. However, while the 
approach is exponential for $\zeta < \zeta_c(g)$, we have 
$\Delta_{\rm OBF} \sim L^{-2}$ for large fields.
For PBF, the system is gapped as for $\zeta < \zeta_c$. However,
in this case corrections are of order $L^{-2}$.

\subsubsection{Intermediate state, $\zeta = \zeta_c(g)$.}

The gap equation is particularly simple. We obtain a solution $\cos k
= 1$ and
\begin{equation}
{\sin (L+1) k\over \sin Lk} = g.
\end{equation}
Hence ${\cal E}_1 = 2 (1 - g)$ with no $L$ dependence. As for the other states
we obtain 
\begin{equation}
{\cal E}_{m+1} = 2 (1 - g) + {g m^2 \pi^2\over 1-g} {1\over L^2} -
    {2 g m^2 \pi^2 \over (1-g)^2} {1\over L^3},
\end{equation}
with $m=1,\ldots L$.
It follows $\Delta_{\rm PBF}
= 4 (1-g) + O(L^{-2})$, as for $\zeta > \zeta_c$. 
However, corrections, of order $L^{-2}$, are a factor of five smaller
compared to the large-field case. For OBF
we obtain the asymptotic behavior
\begin{eqnarray}
\Delta_{\rm OBF} =  {g\over 1-g} \, {\pi^2\over L^2}
- {2g\pi^2 \over (1-g)^2 L^3} + O(L^{-4}).
\label{deltaletac}
\end{eqnarray}
As in the large-field case, $\Delta_{\rm OBF}$ scales as $L^{-2}$. 
However,
the prefactor differs by a factor of 3 
from that obtained for $\zeta>\zeta_c$, cf. Eq.~(\ref{deltalk}).

\subsection{Magnetization and correlation function}
\label{sec6.2}

\begin{figure}[tbp]
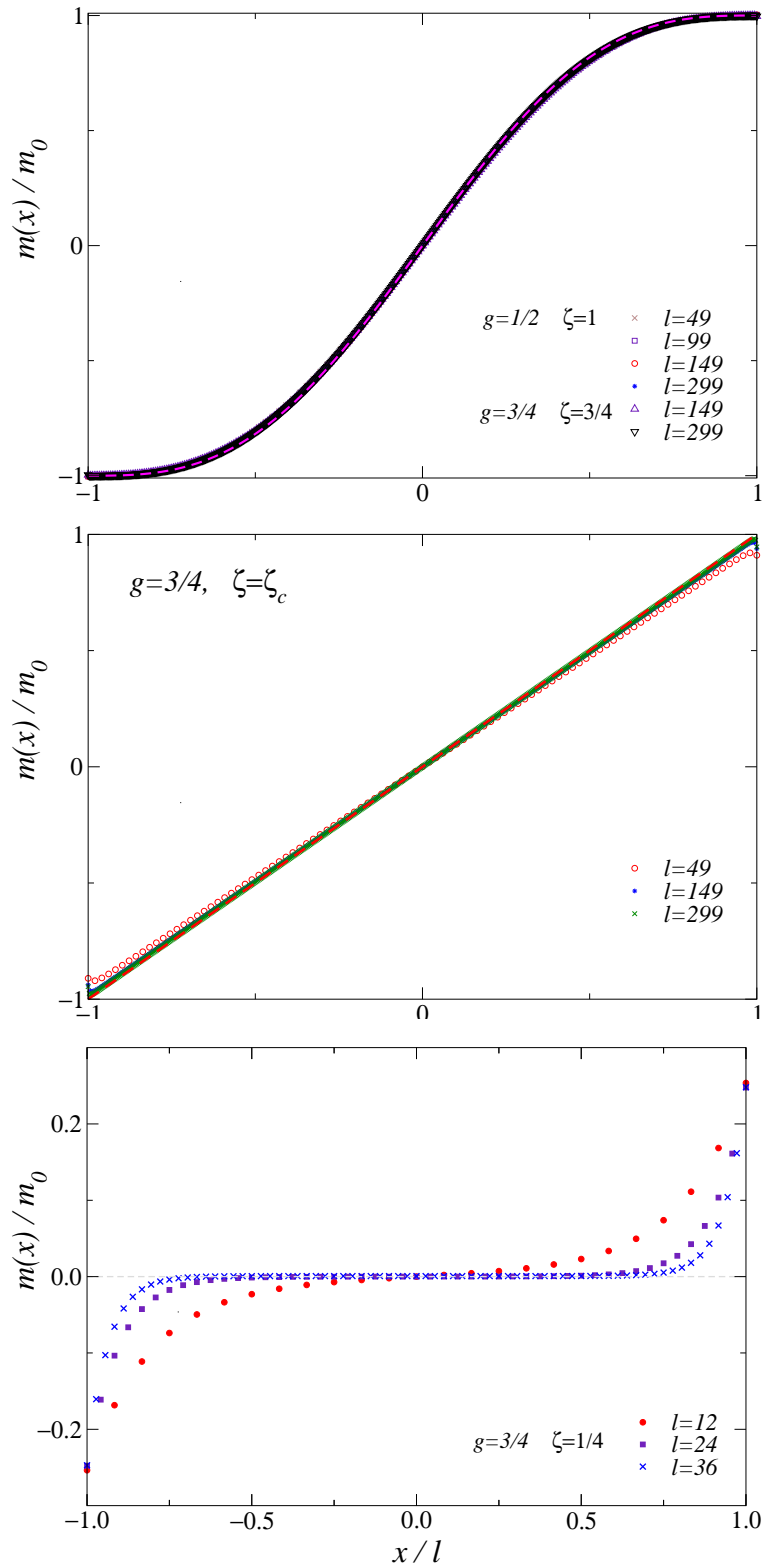

\centerline{\epsfig{width=10truecm,angle=0,file=sxleta.eps}}
\vspace{-2mm}
\centerline{\epsfig{width=10truecm,angle=0,file=sxetac.eps}}
\vspace{-2mm}
\centerline{\epsfig{width=10truecm,angle=0,file=sxseta.eps}}
\caption{Plots of $m(x)/m_0$ versus $x/\ell$ for OBF in the 
  ferromagnetic phase $g < 1$. Here 
  $L = 2 \ell + 1$, $x = 0$ at the center of the chain.
  Top: Results for $g = 1/2$, $\zeta = 1$ and $g = 3/4$, $\zeta = 3/4$
  (large-field kink phase).  Middle: Results for $g =
  3/4$ at $\zeta = \zeta_c = \sqrt{1 - g}$. Bottom: Results for $g =
  3/4$ and $\zeta = 1/4$ (small-field magnetized phase).  
The dashed lines barely visible on top of the data correspond to 
the theoretical curves (\ref{magnetization-scaling}). }
\label{magnetization-fasi}
\end{figure}

We wish now to compute the ground-state magnetization and the
corresponding two-point spin-spin correlation function in the
magnetized phase, i.e., for $ g < 1$. In the PBF case,
the system is magnetized,
hence we expect $m(i)$ and $G(i,j)$ to be both independent of $i$ and
$j$, except possibly close to the boundaries, and to be equal to $m_0$
and $m_0^2$, respectively, where $m_0$ is given in
Eq.~(\ref{modef}). Thus, the boundary magnetic
fields have the only role of lifting the degeneracy.

The OBF case is more interesting. In \ref{App.B} we have fully 
characterized the ground state of the model for small values of $g$.
We find that for it is ferromagnetic for $\zeta < \zeta_c$
(this is consistent with the results for the gap discussed above). 
For $\zeta > \zeta_c$ kink states \cite{Sachdev-book} are the 
relevant low-energy excitations, while for $\zeta = \zeta_c$ the 
ground state is a superposition of kink and ferromagnetic states.
The exact knowledge of the ground state allows us to compute the
ground-state magnetization and the correlation function perturbatively in $g$.
For OBF we find that $m(i) = \langle \sigma^{(1)}_i
\rangle$ vanishes in the low-field case except at the boundaries,
depends linearly on $i$ for $\zeta = \zeta_c(g)$, and varies in a
simple fashion for $\zeta > \zeta_c(g)$. We expect the same space
dependence for all finite values of $g< 1$ in the large-$L$ universal
limit, modulo a multiplicative renormalization. If we set $x = i -
L/2$ and $\ell = L/2$, so that the center of the chain corresponds to
$x=0$, we predict
\begin{equation}
\begin{array}{ll}
 \displaystyle{m(x)\over m_0} = 
  \displaystyle{x\over \ell} + \displaystyle{1\over \pi} 
  \displaystyle \sin {\pi x\over \ell} 
    & \qquad {\zeta > \zeta_c}, \\[2mm]
 \displaystyle{m(x)\over m_0} = \displaystyle{x\over \ell}
    & \qquad {\zeta = \zeta_c}, \\[2mm]
 \displaystyle{m(x)\over m_0} = 0 & \qquad {\zeta < \zeta_c}.
\end{array}
\label{magnetization-scaling}
\end{equation}
These relation should hold in the limit $x\to \infty$, $\ell\to\infty$
at fixed $x/\ell$, except possibly at the boundary, i.e., for $x/\ell
= \pm 1$. Here $m_0$ is the bulk magnetization of the system in the
infinite-volume limit defined in Eq.~(\ref{modef}).
It is interesting to note that the linear behavior of the
magnetization for $\zeta=\zeta_c$ is due to the fact 
that the ground state is translationally invariant: the 
space probability of the kinks is independent of $x$.
Due to the presence of the
boundary fields, the translation invariance of the kink space distribution
is not trivial and, indeed, it does not hold for
$\zeta>\zeta_c$.

To check these predictions and study the approach to the asymptotic
behavior, we perform numerical simulations using the density matrix
renormalization-group (DMRG) method~\cite{DMRG}.
For convenience, we consider chains with
odd $L$, setting $L=2\ell +1 $ (since we are considering ferromagnetic
interactions, the low-energy properties do not depend on this feature,
unlike the antiferromagnetic case, see, e.g.,
Ref.~\cite{LMMS-12}). The results for the magnetization are shown in
Fig.~\ref{magnetization-fasi}. For $\zeta < \zeta_c$, the
magnetization vanishes in a large interval around $x = 0$, assuming
positive and negative values only close to the boundaries. Note that the
region in which $m(x)\not = 0$ shrinks as $L$ increases, conferming
the validity of Eq.~(\ref{magnetization-scaling}) in the large-$L$
limit at fixed $x/\ell$.  For $\zeta > \zeta_c$
we show results for two values of $g$.  Once we renormalize $m(x)$ by
using $m_0$, i.e., we consider the ratio $m(x)/m_0$, results fall one
on top of the other, confirming universality. The results are in full
agreement with the expressions (\ref{magnetization-scaling}).

\begin{figure}[tbp]
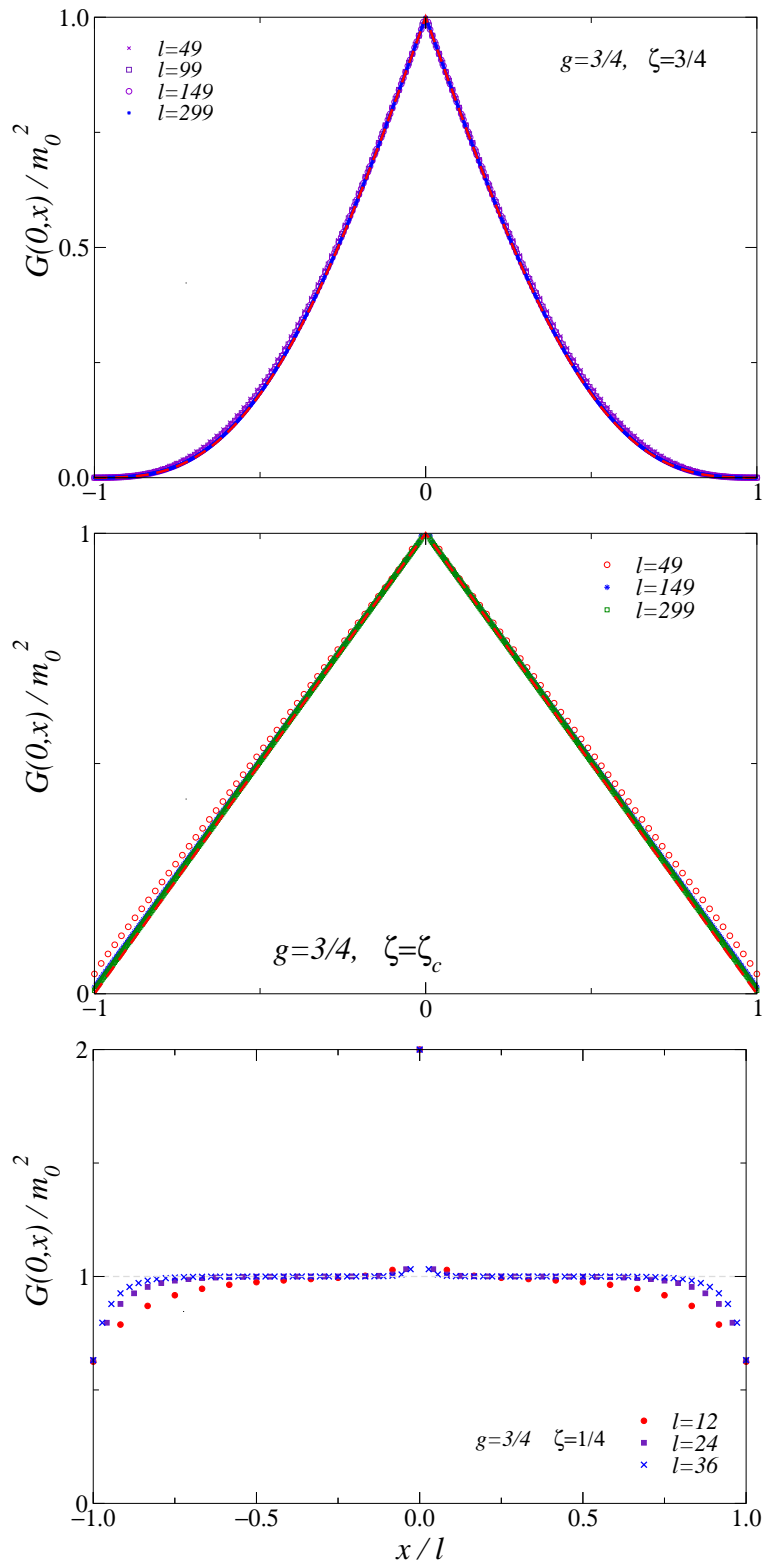

\centerline{\epsfig{width=10truecm,angle=0,file=gxrpi.eps}}
\vspace{-2mm}
\centerline{\epsfig{width=10truecm,angle=0,file=gxetac.eps}}
\vspace{-2mm}
\centerline{\epsfig{width=10truecm,angle=0,file=gxseta.eps}}
\caption{
Plots of $G(y=0,x)/m_0$ versus $x/\ell$ for OBF
($L = 2\ell + 1$, $x = y = 0$ at the center of the domain). 
Top: Results for $g = 3/4$, $\zeta = 3/4$ (large-field kink phase).
Middle: Results for $g = 3/4$ at $\zeta = \zeta_c = \sqrt{1 - g}$.
Bottom: Results for $g = 3/4$ and $\zeta = 1/4$ (small-field magnetized 
phase). 
The dashed lines barely visible on top of the data correspond to 
the theoretical curves obtained by using Eqs.~(\ref{magnetization-scaling})
and (\ref{gij-m}).
}
\label{gx-fasi}
\end{figure}

The same arguments apply to the correlation function $G(i,j)$. Extending 
the small-$g$ results of \ref{App.B} to the whole low-$g$ phase,
we predict 
\begin{equation}
{ G(i,j)\over m_0^2}  = 1 - {1\over m_0} |m(i) - m(j)|
\label{gij-m}
\end{equation}
in the large-$L$ limit. Therefore, the correlation function should be
constant in the magnetized phase, $G(i,j) = m_0^2$, it should depend
linearly on $|i-j|$ for $\zeta = \zeta_c$, while it should have a
sinusoidal dependence in the kink phase. Again, we check these
predictions by comparing them with DMRG data. They are reported in
Fig.~\ref{gx-fasi}. The DMRG data approach the theoretical predictions with
increasing $L$, confirming the theoretical predictions.

\section{Magnet-to-kink transition for $g < 1$}
\label{fktra}

As already discussed in Ref.~\cite{CPV-15}, in the case of OBF
and for $g < 1$, an interesting
universal scaling behavior is observed for $\zeta \to \zeta_c(g)$.
Such a transition is uniquely characterized by the
nature of two phases occurring for $\zeta < \zeta_c$ and $\zeta >
\zeta_c$: on one side the ground state is ferromagnetic, while on the
other side kink states are the relevant low-energy excitations. This
characterization is confirmed by the analysis of the Ising chain on a
ring with a bond defect: at the boundary of the two phases one
observes \cite{CPV-15} a behavior analogous to the one discussed
here. Here, we will report the computation of the energy gap--- the result
was already presented in Ref.~\cite{CPV-15}---and additional results
for the local magnetization, correlation function, and entanglement
entropy. Moreover, we will discuss the density of the fermionic modes
in the equivalent fermionic picture of the model
\cite{Kitaev-01,Alicea-12}, showing that the transition can be
interpreted as a localization transition of the fermionic states at
the boundaries.

The crossover behavior around the transition is parametrized in terms 
of the scaling variable \cite{CPV-15}
\begin{equation}
\zeta_s = { \sqrt{1-g} \over g} \,L \,(\zeta-\zeta_c).
\label{etasdef}
\end{equation}
Then, in the large-$L$ limit we have
\begin{eqnarray}
\Delta_{\rm OBF}(\zeta) \approx \Delta_{\rm OBF}(\zeta_c) f_\Delta (\zeta_s), 
\qquad \xi/L \approx f_\xi(\zeta_s).
\label{sca}
\end{eqnarray}
In general, the scaling functions $f_\Delta (\zeta_s)$ and
$f_\xi(\zeta_s)$ are expected to be universal modulo a $g$-dependent
normalization (which is independent of the observable considered) of
the argument $\zeta_s$. The $g$ dependent prefactor appearing in
Eq.~(\ref{etasdef}) is the required nonuniversal factor. It is determined 
in Sec.~\ref{sec7.1}, analyzing the scaling behavior of the 
gap $\Delta_{\rm OBF}$. According to renormalization-group theory,
this normalization factor is independent of the observable considered, 
hence also the scaling functions of other observables should be 
independent of $g$ apart from a multiplicative factor (not present
if renormalization-group invariant ratios are considered),
 once $\zeta_s$ is used as scaling 
variable. The results we will present for the magnetization and 
the two-point correlation function confirm this prediction.

Beside the gap  $\Delta_{\rm OBF}(\zeta)$, we also  
consider the integrated correlation $\chi$ defined in Eq.~(\ref{chixidef}).
It satisfies the scaling relation
\begin{equation}
 \chi = m_0^2 L f_\chi (\zeta_s),
\label{chisca}
\end{equation}
where $f_\chi (\zeta_s)$ is universal, i.e., $g$ independent.
These results allow us to define the RG dimensions of the different
operators.  In general, we expect
\begin{equation}
\Delta_{\rm OBF}(\zeta) = L^{-z} f_\Delta[L^{y_\zeta} (\zeta-\zeta_c)],\qquad
\chi = A_\chi L^{d - y_\sigma} f_\chi [L^{y_\zeta} (\zeta-\zeta_c)].
\label{scalzetas}
\end{equation} 
Comparison with our results allows us to identify $z = 2$, $y_\zeta =
1$, and $y_\sigma = 0$.  The latter result is consistent with the fact
that a bulk magnetic field coupled to $\sigma^{(1)}$ drives a
first-order quantum transition, see, e.g., Ref.~\cite{CNPV-14}. 

By defining $\zeta_s$ as in Eq.~(\ref{etasdef}) we have implicitly assumed that 
we are fixing $g$ and varying $\zeta$ close to the critical value $\zeta_c(g)$. 
In some contexts, however, it is more natural to consider a 
different point of view. The boundary field strength $\zeta$ is fixed 
(it should satisfy $\zeta < 1$)
and $g$ is varied in a neighborhood of the critical value 
$g_c = 1 - \zeta^2$. In this case we have 
\begin{equation}
\zeta_s = {1\over 2 (1 - \zeta^2)} (g - g_c) L,
\end{equation}
where the $\zeta$ dependent prefactor guarantees that scaling functions are 
the same for all values of $\zeta$.

\begin{figure}[tbp]
\centerline{\epsfig{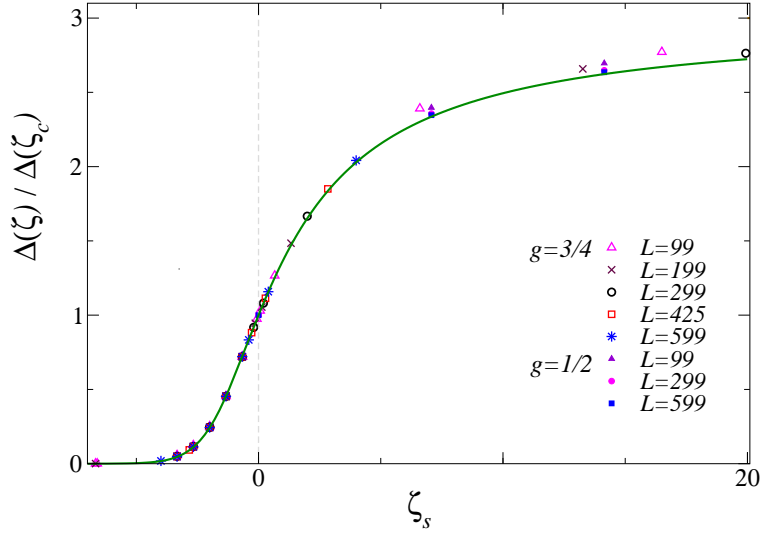}}
%=% \vskip-5mm
\caption{(Color online) Scaling behavior of the gap $\Delta_{\rm OBF}$ 
  around $\zeta=\zeta_c$. The DMRG estimates (points) of 
  $\Delta_{\rm OBF}(\zeta)/\Delta_{\rm OBF}(\zeta_c)$ approach the universal
  theoretical curve for $L\to\infty$ (full line).     }
\label{deetac}
\end{figure}

\subsection{Scaling function for the energy gap} \label{sec7.1}

To determine the scaling behavior of the energy gap $\Delta_{\rm
  OBF}(\zeta)$, we first simplify the secular equation
(\ref{secular-equation}), by taking the limit $\zeta \to \zeta_c$ at
fixed $\zeta_s$. In this limit $k$ scales as $1/L$, so that we set $k
= z/L$.  For $L\to \infty$ the secular equation becomes
\begin{equation}
4 \zeta_s z + (4 \zeta_s^2 - z^2) \tan z = 0.
\label{sec-etac-1}
\end{equation}
Expressing $\tan z$ as a function of $\tan z/2$, we find that the
solutions of Eq.~(\ref{sec-etac-1}) satisfy one of the two equations
\begin{eqnarray}
&&\tan {z\over2} = - {z\over 2\zeta_s}, \label{eq-tanz2-1} \\
&& \tan {z\over2} = {2\zeta_s \over z}. \label{eq-tanz2-2} 
\end{eqnarray}
If $z_{a1} < z_{a2} < \ldots$ and $z_{b1} < z_{b2} < z_{b3}\ldots$ are
the positive solutions of Eqs.~(\ref{eq-tanz2-1}) and
(\ref{eq-tanz2-2}), respectively, it is easy to verify that $z_{b1} <
z_{a1} < z_{b2} < \ldots$ for $\zeta_s \ge 0$ and $\zeta_s \le
-1$. For $-1\le \zeta_s < 0$ we have instead $z_{a1} < z_{b1} < z_{a2}
\ldots$. For $\zeta_s < 0$ there are also localized solutions with $k
= i u/L$. The parameter $u$ satisfies one of the two equations
\begin{eqnarray}
&& \tanh {u\over2} = - {u\over 2\zeta_s}, \label{eq-tanw2-1} \\
&& \tanh {u\over2} = - {2\zeta_s \over u}. \label{eq-tanw2-2} 
\end{eqnarray}
Eq.~(\ref{eq-tanw2-2}) has a positive solution $u_b$ for $\zeta_s < 0$, while 
Eq.~(\ref{eq-tanw2-1}) has a positive solution $u_a$ satisfying $u_a< u_b$ only
for $\zeta_s < -1$. These results allow us to determine the scaling 
function defined in Eq.~(\ref{sca}). We obtain 
$f_\Delta(\zeta_s) = (z_{a1}^2 - z_{b1}^2)/\pi^2$ for $\zeta_s > 0$,
$f_\Delta(\zeta_s) = (z_{a1}^2 + u_{b}^2)/\pi^2$  for $-1 < \zeta_s < 0$ 
and $f_\Delta(\zeta_s) = (u_b^2 - u_a^2)/\pi^2$ for $\zeta_s < -1$.

We can easily determine the asymptotic behaviors. 
For small $\zeta_s$ we have 
\begin{equation}
f_\Delta(\zeta_s)= 1 + {4\zeta_s\over \pi^2} + O(\zeta_s^2),
\end{equation}
while for $\zeta_s \to \infty$ we have 
\begin{equation}
f_\Delta(\zeta_s)= 3 - {6\over \zeta_s} + O(\zeta_s^{-2}).
\end{equation}
Finally, for $\zeta_s \to -\infty$ we have 
\begin{equation}
f_\Delta(\zeta_s)\approx {32\over \pi^2} \zeta_s^2 e^{2\zeta_s}.
\end{equation}
The theoretical curve is shown in Fig.~\ref{deetac}.  The DMRG data
confirm the theoretical calculation: the numerical estimates of the ratio
$\Delta_{\rm OBF}(\zeta)/\Delta_{\rm OBF}(\zeta_c)$ clearly approach
the scaling curve $f_\Delta(\zeta_s)$ with increasing $L$.

\subsection{Magnetization and two-point function}

\begin{figure}[tbp]
\centerline{\epsfig{width=10truecm,angle=-90,file=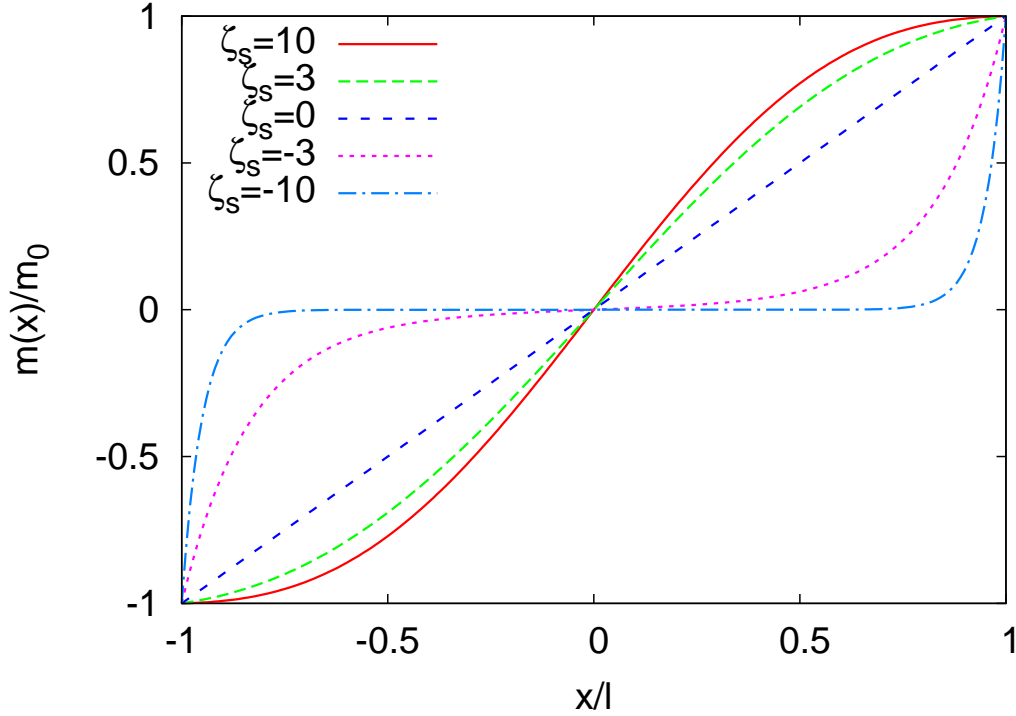}}
%=% \vskip-5mm
\caption{(Color online) Rescaled local magnetization $m(x)/m_0$ versus 
  $x/\ell$ for $\zeta_s = -10,-3,0,3,10$.
}
  \label{magnxcr}
\end{figure}

\begin{figure}[tbp]
\centerline{\epsfig{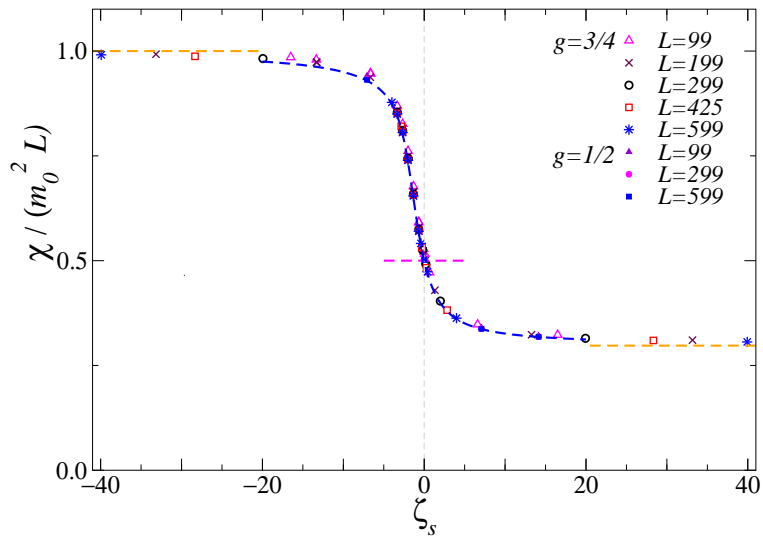}}
%=% \vskip-5mm
\caption{(Color online) Crossover scaling curve for $\chi/(m_0^2 L)$
versus $\zeta_s$.
  We report the theoretical prediction (dashed line)
  and DMRG data for $g=1/2$ and $g=3/4$
  to confirm universality. 
  Note the limiting values:
  $f_\chi(0)=1/2$, $f_\chi(-\infty)=1$, and $f_\chi(\infty)\approx 0.297$,
  indicated by the horizontal dashed lines.  
}
\label{chietacr}
\end{figure}

\begin{figure}[tbp]
\centerline{\epsfig{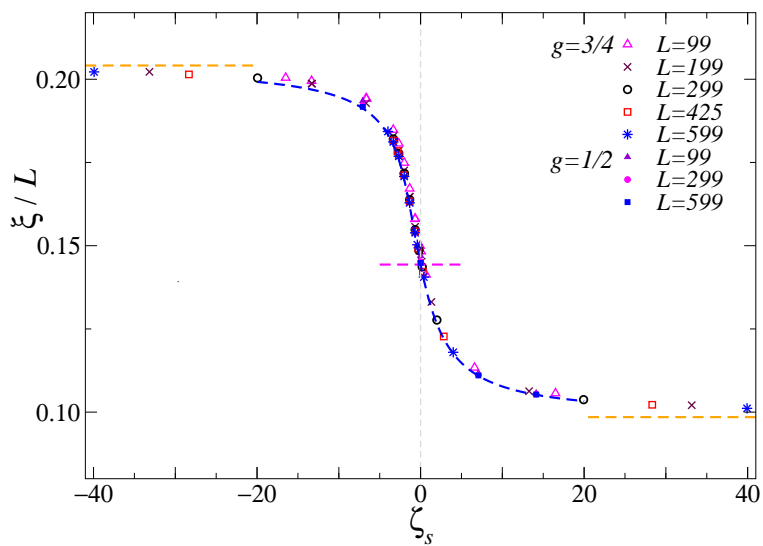}}
%=% \vskip-5mm
\caption{(Color online) Crossover scaling curve $\xi/L = f_\xi(\zeta_s)$ 
  for the correlation length $\xi$. We report the theoretical prediction for
  the asymptotic scaling behavior and DMRG data for $g=1/2$ and $g=3/4$ to
  confirm universality.  Note the limiting values:
  $f_\xi(0)=1/\sqrt{48}$, $f_\xi(-\infty) = 1/\sqrt{24}$, and
  $f_\xi(\infty)\approx 0.0985$.  }
  \label{xietacr}
\end{figure}

Let us now consider the behavior of the correlations close to
$\zeta_c$.  The local magnetization $m(i)$ and the correlation function
$G(i,j)$ are computed exactly in the limit $g \to 0$ in
\ref{App.B4}. As we have already done in Sec.~\ref{sec6.2}, we extend
these expressions to all values of $g$ satisfying $g<1$, by replacing
$m(i)$ with $m(i)/m_0$ and $G(i,j)$ with $G(i,j)/m_0^2$.  Therefore,
we predict for $\zeta_s > 0$
\begin{equation}
{m(x) \over m_0} = {z\over z + \sin z} \left [{1\over z} \sin \left({ z x \over
\ell} \right) + {x\over \ell} \right],
\end{equation}
where $z$ is the smallest positive solution ($z=z_{b1}$) of 
Eq.~(\ref{eq-tanz2-2}), $x = i - L/2$, and $\ell = L/2$, as before. 
For $\zeta_s < 0$ we have instead 
\begin{equation}
{m(x) \over m_0} = {u\over u + \sinh u} 
\left [{1\over u} \sinh \left({ u x\over
\ell} \right) + {x\over \ell} \right],
\end{equation}
where $u$ is the solution ($u = u_b$) of Eq.~(\ref{eq-tanw2-2}). 
As for the correlation function $G(i,j)$
it satisfies Eq.~(\ref{gij-m}). 
It is important to stress
that the previous expressions, although derived for small values of $g$, 
are expected to be {\em exact} for any $g < 1$. For instance, since 
$m(x)/m_0$ is a dimensionless renormalization-group invariant ratio, 
it should scale as 
\begin{equation}
m(x,\zeta,g,L) = m_0 f_{m}[a(g) (\zeta - \zeta_c) L, x/L],
\end{equation}
where $f_m$ is a universal, hence $g$ independent, function. The specific
features of the model enter in this expression only through the 
normalization nonuniversal factor $a(g)$ and the critical value $\zeta_c(g)$. 
The factor $a(g)$ does not
depend on the observable, hence it can be determined by using any quantity. 
We use the gap $\Delta_{\rm OBF}$, 
for which we are able to obtain exact results 
for any $g < 1$, defining $\zeta_s = a(g) (\zeta - \zeta_c) L$.
Therefore, $f_m$ as a function of $\zeta_s$ and $x/L$ is $g$ independent. 
Therefore, our results, obtained in the limit $g\to 0$, hold for any value
of $g$ satisfying $g < 1$.

The previous expressions for $m(x)$
hold in the limit $x\to \infty$, $L\to \infty$ at fixed $x/\ell$ and
smoothly interpolate between the expressions valid in the different
phases.  Note that we have $m(i) \to \pm m_0$ for $x/\ell\to\pm 1$ in
all cases.  This result does not however necessarily apply to the
boundary points, i.e., for $x/\ell = \pm 1$, as the magnetization is
not continuous at the boundary.  For instance, in the magnetized
phase, $m_i = 0$ in all interior points, while $m_1$ and $m_L=-m_1$
are nonuniversal functions of the field strength $\zeta$.

A graph of $m(x)/m_0$ for several values of $\zeta_s$ as a function of
$x/\ell$ is reported in Fig.~\ref{magnxcr}. As expected, for $\zeta_s
\to -\infty$ the magnetization approaches zero exponentially in an
interval that is centered in the middle of the chain and that widens
as $|\zeta_s|$ increases.  More precisely, for $\zeta_s \to -\infty$ we
have $u \approx - 2\zeta_s$ and
\begin{equation}
m(x) = {\sinh 2 \zeta_s x/\ell \over \sinh 2\zeta_s},
\end{equation}
which shows that $m(x)$ vanishes except in two tiny intervals close to
the boundaries of width $\ell/(2 |\zeta_s|)$.

To verify the prediction (\ref{gij-m}) for $G(i,j)$ we compute the
scaling functions associated with $\chi$ and $\xi$. They are reported
in \ref{App.B4} and compared with DMRG data in Figs.~\ref{chietacr}
and \ref{xietacr}. 
Results with different values of 
$g$ fall on top of the theoretical curve when plotted 
versus $\zeta_s$. This confirms our predictions for $G(i,j)$ and the 
magnetization, as well as the correctness of the nonuniversal prefactor 
appearing in the definition of $\zeta_s$.

\subsection{Equivalent  fermionic picture}

It is interesting to reinterpret our results in the equivalent
fermionic picture of the model.  In the ferromagnetic phase, i.e., for
$\zeta<\zeta_c$, the lowest eigenstates are superpositions of Majorana
fermionic states localized at the boundaries
\cite{Kitaev-01,Alicea-12}.  In finite systems, their overlap does not
vanish, giving rise to the splitting $\Delta \sim e^{-L/l_0}$.  The
coherence length $l_0$ diverges at the kink-to-magnet transitions as
$l_0^{-1}\sim |\ln s|\sim \zeta_c-\zeta$, a behavior analogous to that
observed at the order-disorder transition $g\to 1^-$ where
$l_0^{-1}\sim |\ln g|$.

\begin{figure}[tbp]
\centerline{\epsfig{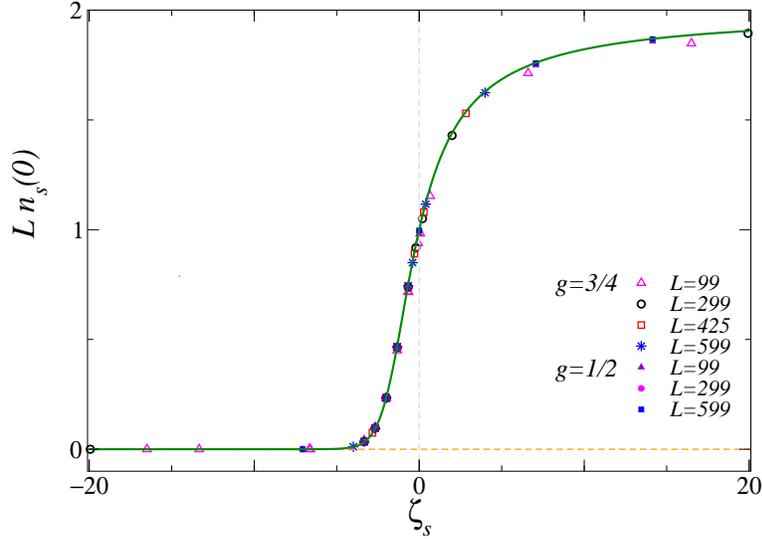}}
%=% \vskip-5mm
\caption{(Color online) Scaling of the subtracted 
  fermionic density at the center
  of the chain. We plot DMRG data of $Ln_s(0)$ for $g=3/4$ and $g=1/2$ versus
  $\zeta_s$.  The full line corresponds the asymptotic analytic curve,
  cf. Eqs.~(\ref{ns-eq1}) and (\ref{ns-eq2}).
  Note that $Ln_s(0)$ approaches 2 for
  $\zeta_s \to \infty$, 0 for $\zeta_s \to -\infty$, while it is equal
  to 1 for $\zeta_s = 0$.  }
\label{fig:Deltan0}
\end{figure}

Within this picture, it is natural to consider
\begin{equation}
n(i) = {1 + \langle \sigma_i^{(3)}\rangle \over 2},
\label{nxdef}
\end{equation} 
which can interpreted as the particle density in the fermionic
representation of the Ising chain, see Eq.~(\ref{sigma-c-relation}). 
Its large-$L$ limit is given by~\cite{Pfeuty-70}
\begin{equation}
n_h = {1\over 2} + {1\over 2\pi}\int_0^\pi
dk\,{g+{\rm cos}k\over \sqrt{1 + g^2 + 2g \,{\rm cos}k}}.
\label{n0}
\end{equation}
We consider the difference 
\begin{equation}
n_s(x) \equiv n(i) - n_h,
\label{defden}
\end{equation}
where $x = i - L/2$, as before.  Such a quantity has been computed for
small values of $g$ in \ref{App.B4}.  The universality argument we
have used for $m(i)$ and $G(i,j)$ can be applied to $n_s(x)$. We
therefore predict
\begin{equation}
L n_s(x) = {z\over z + \sin z} \left (1 + \cos {z x\over \ell}\right)
\label{ns-eq1}
\end{equation}
for $\zeta_s > 0$ and 
\begin{equation}
L n_s(x) = {u\over u + \sinh u} \left (1 + \cosh {u x\over \ell}\right)
\label{ns-eq2}
\end{equation}
for $\zeta_s < 0$, where $z$ and $u$ are determined as in the case of
the local magnetization. To verify this prediction we compute $n_s(0)$
numerically for $g = 1/2$ and $g = 3/4$.  The numerical results are
compared with theory in Fig.~\ref{fig:Deltan0}. The agreement confirms
the analytic prediction.

\begin{figure}[tbp]
\centerline{\epsfig{width=10truecm,angle=-90,file=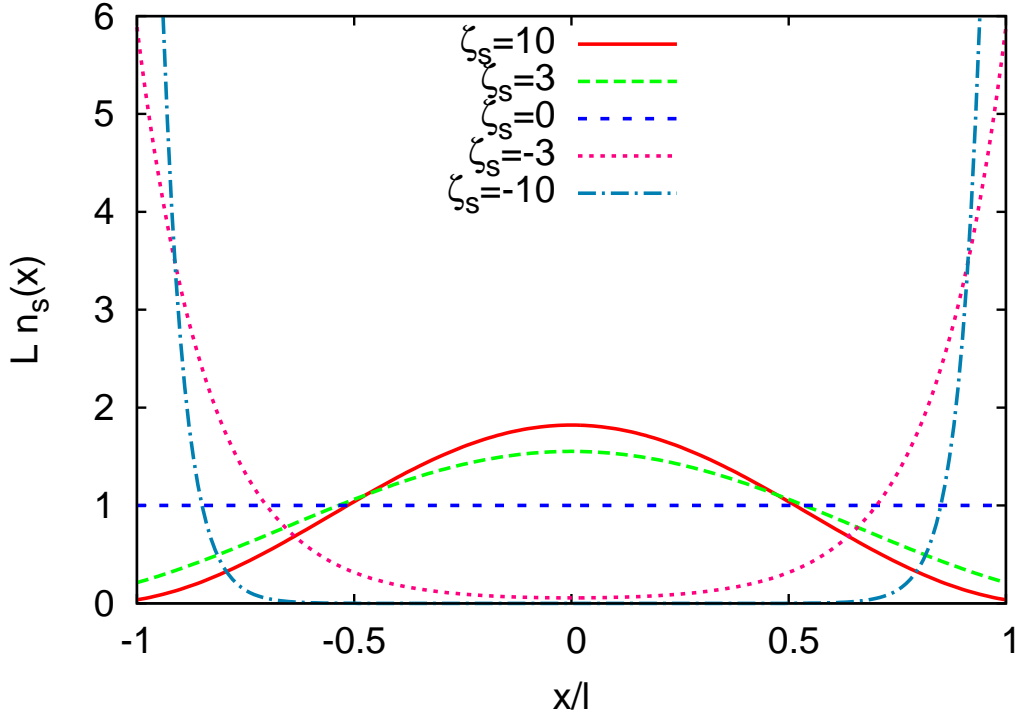}}
%=% \vskip-5mm
\caption{(Color online) Subtracted fermionic density $L n_s(x)$ versus 
  $x/\ell$ for $\zeta_s = -10,-3,0,3,10$ ($x = 0$ at the center of the chain).
}
  \label{Deltanxcr}
\end{figure}

\begin{figure}[tbp]
\centerline{\epsfig{width=10truecm,angle=-90,file=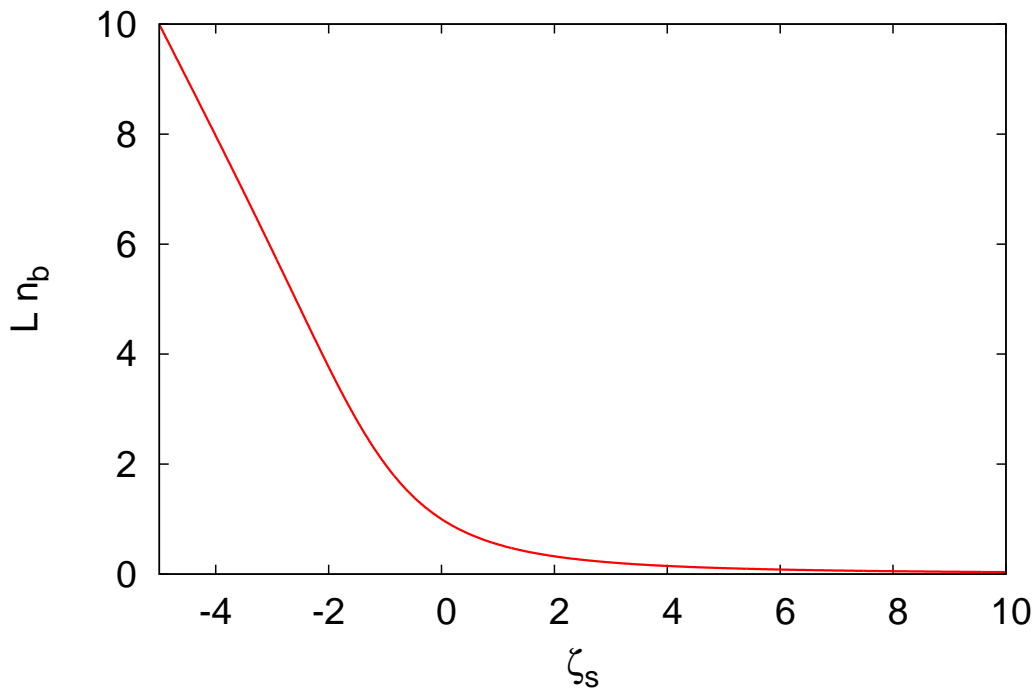}}
%=% \vskip-5mm
\caption{(Color online) Subtracted fermionic density $ L  n_b = 
  L n_s(\pm \ell)$ at the boundaries of the chain versus $\zeta_s$. 
  We have $L n_b (\zeta_s) \approx \pi^2/(2 \zeta_s^2)$ for 
  $\zeta_s\to \infty$, 
  $L n_b (\zeta_s) \approx - 2 \zeta_s$ for 
  $\zeta_s\to -\infty$, and $L n_b (0) = 1$.
}
\label{Deltanb}
\end{figure}

The subtracted fermionic density $L n_s(x)$ is
reported in Fig.~\ref{Deltanxcr} as a function of $x/\ell$.  For
positive values of $\zeta_s$ it has a maximum at the origin and shows
that fermions are delocalized.  For $\zeta_s = 0$, $L n_s(0)$ is
constant, while for negative values of $\zeta_s$, there is a
significant enhancement at the boundary.  For $\zeta_s \to -\infty$,
as in the case of the magnetization, $L n_s(x)$ is different from
0 only close to the boundary, in an interval of width
$\ell/2|\zeta_s|$. It is interesting to consider the behavior of $L
n_s(x)$ at the boundaries, i.e., for $x\to \pm \ell$, which is shown
in Fig.~\ref{Deltanb}.  In the kink phase, such a quantity scales as
$\pi^2/(2 \zeta_s^2)$ for $\zeta_s \to \infty$. On the other hand it
diverges as $- 2\zeta_s$ for $\zeta_s \to -\infty$, that is when the
magnetized phase is approached.  As such, it represents a physically
relevant order parameter for the transition, which distinguishes the
two different phases.

\subsection{Half-chain entanglement entropy}

\begin{figure}[tbp]
\centerline{\epsfig{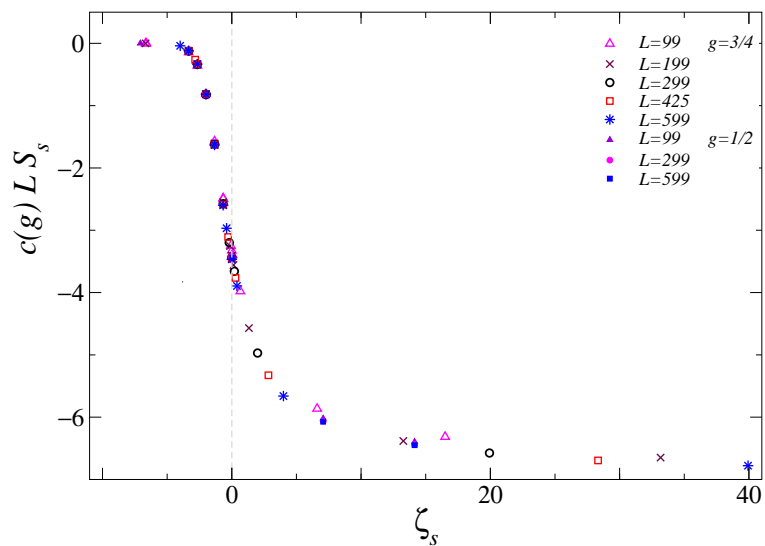}}
%=% \vskip-5mm
\caption{(Color online) Scaling of the subtracted half-chain
  entanglement entropy. The DMRG data appear to approach an asymptotic
  scaling curve. The scaling curves for different values of $g$
  match after a global nonuniversal rescaling: fixing $c(g=3/4)=1$
  the optimal matching is obtained for $c(g=1/2)\approx 1.85$.
  }
\label{entaen}
\end{figure}

Finally, we discuss the behavior of the half-chain entanglement
entropy~\cite{CCD-09}. We divide the chain into two connected parts
$[-\ell,0]$ and $[1,\ell]$ and consider the von Neumann entropy
\begin{equation}
S = -{\rm Tr}\,\rho_{[-\ell,0]} \ln \rho_{[-\ell,0]},
\label{vNen}
\end{equation} 
where $\rho_{[-\ell,0]}$ is the reduced density matrix of the
subsystem $[-\ell,0]$, i.e.,
\begin{equation}
\rho_{[-\ell,0]}={\rm Tr}_{[1,\ell]} \;\rho ,
\end{equation}
and $\rho$ is the density matrix of the ground state.  The
entanglement entropy $S$ approaches a constant in the large-$L$ limit,
except for the critical case $g=1$ where it increases
logarithmically~\cite{CCD-09}. Its large-$L$ limit for OBC is given
by~\cite{Peschel-04}
\begin{eqnarray}
&&S_{\rm OBC}(g) = \ln 2 
+ {1\over 12}
\left[ \ln{g^2\over 16\sqrt{1-g^2}} + \left(1 - {g^2\over 2}\right)
{4 I(g) I(\sqrt{1-g^2})\over \pi}\right],
\nonumber\\
&&I(z) = \int_0^1 {dx\over \sqrt{(1-x^2)(1-z^2 x^2)}} .
\label{sobc}
\end{eqnarray}
In Fig.~\ref{entaen} we plot the difference
\begin{equation}
S_s \equiv S(\zeta,g,L) - S_{\rm OBC}(g),
\label{dsdef}
\end{equation}
which shows the scaling behavior
\begin{equation}
S_s   \approx L^{-1} f_S(\zeta_s), 
\label{scashc}
\end{equation}
where $f_S$ is a universal function apart from a 
multiplicative nonuniversal constant. 
These results imply that the large-$L$ limit
of the half-chain entanglement entropy does not depend on the boundary
conditions, being the same
for OBC and OBF.  However, the
$O(L^{-1})$ corrections are affected by the boundary term (\ref{hb}),
giving rise to a nontrivial universal behavior around $\zeta_c$.

\subsection{Comparison with the existing results for two-dimensional classical
models} \label{comp-2D}

The results we have obtained allow us to derive exact scaling functions 
for the classical two-dimensional Ising model in a strip 
$L\times \infty$ in the presence of two equal magnetic fields on the 
boundaries. This model has been extensively studied, especially in the 
case of opposite magnetic fields. In this case, the wetting 
(or interface localization-delocalization) transition,
which is the analog of the magnet-to-kink transition we are considering 
here, has been extensively studied 
\cite{Abraham-80,NF-82,Ciach-86,CS-87,PS-88,PE-90,PEN-91,SMO-94,%% 
MS-96,Maciolek-96}. We can therefore check some of our results, comparing 
them  with the existing ones in the literature.

The results for the gap $\Delta_{\rm OBF}$ obtained in Sec.~\ref{sec7.1} 
can be used to derive the scaling behavior of the longitudinal 
correlation length $\xi_\Vert$, defined by $\xi_\Vert = 1/\Delta_{\rm OBF}$. 
Such a quantity was computed in the restrictive solid-on-solid model in 
Ref.~\cite{PS-88}. Their result is completely consistent with ours. 
If our scaling quantities $z$ and $u$ defined in Sec.~\ref{sec7.1} 
are identified with the quantities $\sqrt{2 G_0}$, $\sqrt{2 G_1}$
appearing in
Ref.~\cite{PS-88}, we obtain exactly the same expression for $\xi_\Vert/L^2$.

Several papers study the magnetization profile, which gives information 
on the behavior of the interface between the two coexisting phases.
The behavior in the large-field phase was determined in Ref.~\cite{Ciach-86}
by using the restrictive solid-on-solid model. The result is in agreement
with our prediction (\ref{magnetization-scaling}) for $\zeta > \zeta_c$. 
The linear behavior observed at $\zeta = \zeta_c$ was also derived 
at the wetting transition
by using the solid-on-solid model \cite{PEN-91}. No exact results are 
instead available for the magnetization profiles in the whole crossover 
region, which, in the classical case, is usually parametrized in terms of
$[T - T_w(h)]L$, where $h$ is the boundary field and 
$T_w$ the corresponding wetting temperature (in the quantum-to-classical
mapping $g$ corresponds to $T$, hence this scaling corresponds to varying 
$g$ at fixed $\zeta$). 
However, the numerical results for the magnetization reported in 
Refs.~\cite{SMO-94,MS-96,Maciolek-96} (and, in particular, Fig.~2 of 
Ref.~\cite{SMO-94}) appear to be in full agreement with our 
prediction shown in Fig.~\ref{magnxcr}. 

\section{Conclusions}
\label{conclu}

We consider the one-dimensional quantum Ising chain in the presence 
of a transverse magnetic field $g$ \cite{Pfeuty-70} and of boundary 
magnetic fields aligned with the order-parameter spin operator. 
In particular, we consider fields
with equal strength $\zeta$, which have either the same (PBF) or the
opposite (OBF) direction, cf. Eqs.~(\ref{pbfdef}) and (\ref{obfdef}).
We assume the chain to have length $L$ and derive the finite-size 
behavior of several quantities in the limit $L\to \infty$ as a function
of $\zeta$.

We derive analytic predictions for the gap in all phases as a function of the 
boundary field strength. In the paramagnetic quantum phase, the 
leading behavior is independent of the boundary conditions. At the critical
transition $g = 1$, finite-size scaling depends on the boundary fields. 
The point $\zeta = 0$, corresponding to open boundary conditions, is 
a surface critical point. We derive analytic expressions for the 
scaling functions that parametrize the crossover behavior for $g\to 1$
and any $\zeta$ (bulk critical behavior), and for $g = 1$
and $\zeta \to 0$ (surface critical behavior). 
In the quantum ferromagnetic phase $g < 1$, 
if the boundary fields are oppositely aligned, 
the finite-size behavior drastically changes as 
$\zeta$ varies \cite{CPV-15}. For small $\zeta$, the system is in a 
ferromagnetic phase in which the gap decreases exponentially with the system
size. On the other hand, for large fields, kink propagating modes of 
momenta $1/L$ are the relevant low-energy excitations, so that the 
gap decreases as $1/L^2$. The two phases are separated by a critical transition
at $\zeta = \zeta_c(g)$.
Close to this transition,
low-energy properties show a universal scaling behavior in terms of 
the scaling variable $L(\zeta-\zeta_c)$. The transition 
is only characterized by the nature of the coexisting phases.
Indeed, the same transition occurs
in Ising rings in the presence of a localized bond defect \cite{CPV-15}.

It is interesting to interpret our results in the context of two-dimensional
classical Ising models, using the quantum-to-classical mapping. 
As is well known \cite{Sachdev-book}, the quantum Ising chain of length
$L$ corresponds to a classical Ising model on a strip of width $L$, the 
transverse field playing the role of the temperature. Therefore,
our results provide exact predictions for the Ising model in an 
$L\times\infty$ geometry with boundary fields. Our results for the 
gap at the critical point can be used to predict the FSS behavior of the 
longitudinal exponential correlation function $\xi_\Vert= 1/\Delta$ as 
a function of $w = A (T-T_c) L$ and of  $\zeta_b = B \zeta L^{1/2}$, 
where $A$ and $B$ are nonuniversal model-dependent constants. Moreover,
our extensive results at the magnet-to-kink transition for $g < 1$  for OBF
provide predictions for the finite-size scaling functions 
at the classical wetting transition (sometimes also 
called interface localization/delocalization 
transition \cite{BLM-03}) in two dimensions. 
In particular, we have exact predictions for the FSS behavior of the 
magnetization profile and of the correlation function of the layer magnetization
in the whole crossover region close to the transition.

In the context of the wetting transition, 
one may also consider a second important surface parameter, 
which is usually referred to
as surface enhancement \cite{NF-82,Dietrich-88}. While a
boundary magnetic field is the most relevant magnetic 
perturbation that breaks the 
${\mathbb Z}_2$ symmetry at the boundary, surface enhancement is the 
most relevant (energy-like) perturbation that is ${\mathbb Z}_2$ invariant.
In the context of the 
quantum chain, such an interaction can be mimicked by adding a 
term ($h_1 \sigma_1^{(3)} + h_L \sigma_L^{(3)})$, localized at the boundary, 
to the general Ising-chain Hamiltonian.
It is not difficult to generalize the results presented here to this more 
general case (results for the spectrum in the absence of boundary magnetic
fields appear in Refs.~\cite{BS-74,BG-87}). 
At the critical point, this would allow us to study 
the behavior at the ordinary and extraordinary surface transitions 
(in the absence of surface enhancement the point $g = 1$, $\zeta = 0$ 
corresponds to the so-called special transition \cite{NF-82}). Moreover,
a complete understanding of the phase behavior for $g < 1$ in 
the absence of a bulk magnetic field should be possible. 
In the quantum case, these results would also 
be relevant for Ising quantum rings with localized site defects, a distinct 
case with respect to that considered in Ref.~\cite{CPV-15}.

\appendix

\section{Spectrum determination} \label{App.A}

\subsection{Preliminary matrix results} \label{App.A1}

In this appendix we collect results on the spectrum 
of some matrices that are relevant for the discussion.  We
begin by considering the $n\times n$ matrix (we write it for $n=5$)
\begin{equation}
A_n(d,b) = \begin{pmatrix} 
   d  & b & 0 & 0 & 0 \\
   b  & d & b & 0 & 0 \\
   0  & b & d & b & 0 \\
   0  & 0 & b & d & b \\
   0  & 0 & 0 & b & d 
\end{pmatrix} \; .
\end{equation}
To determine its eigenvalues, we compute 
\begin{equation} 
   a_n(d,b) = \hbox{det}\, (A_n(d,b) - \lambda I).
\end{equation}
If we set $\lambda = d - 2 b \cos k$, the determinant $a_n(d,b)$ satisfies the 
recursion relation
\begin{equation}
   a_n = 2 b \cos k a_{n-1} - b^2 a_{n-2}.
\label{recursion-an}
\end{equation}
The solution is 
\begin{equation}
a_n b^{-n} = c_1 e^{ikn} + c_2 e^{-ikn},
\end{equation}
where $c_1$ and $c_2$ are arbitrary constant. If we require 
$a_1 = 2 b \cos k$ and $a_2 = b^2 (4 \cos^2 k - 1)$, we obtain 
\begin{equation}
c_1 = - {i e^{ik}\over 2 \sin k}, \qquad c_2 = c_1^*.
\end{equation}
It follows 
\begin{equation}
a_n = b^n {\sin k (n+1)\over \sin k}.
\end{equation}
The eigenvalues are solutions of the equation
$a_n = 0$. We obtain $n$ eigenvalues given by
\begin{equation}
\lambda_m = d - 2 b \cos k_m \qquad k_m = {\pi m \over n + 1} ,
\end{equation}
with $m = 1,\ldots n$. 

Let us now compute the eigenvectors. Let $(\alpha_1,\ldots,\alpha_n)$ be 
the eigenvector corresponding to eigenvalue $\lambda_m$. 
It satisfies the relations
\begin{eqnarray}
&&  2 \alpha_1 \cos k_m + \alpha_2 = 0, \\
&&  \alpha_{j-1} + 2 \alpha_j \cos k_m + \alpha_{j+1} = 0, \qquad 
j = 2,\ldots n - 1 
\label{eigenvectors-relation-2} \\
&&  \alpha_{n-1} + 2 \alpha_n \cos k_m = 0.
\end{eqnarray}
The solution of the recursion relation 
(\ref{eigenvectors-relation-2}) is 
\begin{equation}
\alpha_j = (-1)^j (e_1 e^{ik_m j} + e_2 e^{-ik_m j}),
\end{equation}
where $e_1$ and $e_2$ are two arbitrary constants.
Requiring $\alpha_2 = - 2 \cos k_m \alpha_1$, we obtain $e_1 + e_2 = 0$, 
so that 
\begin{equation}
\alpha_j = (-1)^j e_1 \sin k_m j,
\end{equation}
where we have rescaled $e_1$ for convenience. Requiring the eigenvector to have 
unit norm, we obtain 
\begin{equation}
\alpha_j = (-1)^j \sqrt{2 \over n+1} \sin k_m j.
\end{equation}
We also determine the eigenvalues of the 
matrix
\begin{equation}
{\tilde A}_n(d,b,e) = \begin{pmatrix} 
   e  & b & 0 & 0 & 0 \\
   b  & d & b & 0 & 0 \\
   0  & b & d & b & 0 \\
   0  & 0 & b & d & b \\
   0  & 0 & 0 & b & e 
\end{pmatrix} \; ,
\label{matrix-Atilde-def}
\end{equation}
which differs from $A$ only for two matrix elements: ${\tilde A}_{n,11}$ 
and ${\tilde A}_{n,nn}$ are equal to $e$. Setting again 
\begin{equation}
{\tilde a}_n = {\rm det}\, ({\tilde A}_n - \lambda I),
\end{equation}
and $\lambda = d - 2 b \cos k$, we obtain 
\begin{equation}
{\tilde a}_n = (e - d + 2 b \cos k)^2 a_{n-2} - 
       2 b^2 (e - d + 2 b \cos k) a_{n-3} + b^4 a_{n-4}.
\end{equation}
Using the recursion relation (\ref{recursion-an}), we obtain finally
\begin{equation}
{\tilde a}_n = (1 - \delta^2) a_n + 
             2 b \delta (\delta \cos k - 1) a_{n-1},
\label{recursion-tildean}
\end{equation}
where $\delta = (d-e)/b$.

\subsection{Secular equation}

To determine the spectrum of Hamiltonian (\ref{def-Hextended}), 
we compute the 
eigenvalues of the matrix $C$ defined in Eq.~(\ref{matrixC-def}).
We write them as 
\begin{equation}
\lambda = g^2 + 1 - 2 g \cos k.
\end{equation}
Note that nothing forbids $\lambda$ to be larger than $|1+g|$ or
smaller that $|1-g|$, hence $\cos k$ is not necessarily  
bounded between $-1$ and $1$. Hence, 
we must consider real values of $k$ with $0\le k\le \pi$
and complex values of the form $k = ih $ and $k = \pi + ih$ with $h > 0$.

We are not interested in the zero eigenvalue of $C$, hence we will only 
consider
the $(L+1)\times (L+1)$ matrix formed by the first $(L+1)$ rows and columns,
which we label with the symbol $C_{L+1}$. Then, we introduce a matrix 
$D_{L+1}$ defined as (here $L=4$)
\begin{equation}
D_{L+1} = \begin{pmatrix}
    g^2 + 1   &   g    &     0     &      0      &     0   \\
     g       &  g^2 + 1 & g &      0      &     0   \\
      0         &     g      & g^2 + 1 & g   &     0   \\
      0         &     0      & g      & g^2 + 1 & g    \\
      0         &     0      &     0     &      g & g^2 + 1
     \end{pmatrix}
\end{equation}
The corresponding secular equations are
\begin{eqnarray}
&& c_n = \hbox{det}\, (C_n - \lambda I) \qquad 
d_n = \hbox{det}\, (D_n - \lambda I).
\end{eqnarray}
The determinant $d_n$ has already been computed in 
\ref{App.A1}: 
\begin{equation}
d_n = g^n {\sin k (n+1) \over \sin k}.
\label{dn-result}
\end{equation}
To determine $c_n$, we expand the determinant with respect to the
first and the last rows. We obtain the following recursion relation:
\begin{eqnarray}
c_n &=&
 (g^2 + 1 - J_0^2 - 2 g \cos k)(1 - J_L^2 - 2 g \cos k)
 d_{n-2}  \\
&& +
 g^2 (g^2 + 1 - J_0^2 J_L^2 - 2 g \cos k - 2 g J_0^2 \cos k)
 d_{n-3} + g^4 J_0^2 d_{n-4}.\nonumber 
\end{eqnarray}
A simpler expression is obtain by eliminating $d_{n-4}$ and $d_{n-3}$,
and expressing the result in terms of $d_{n-1}$ and $d_{n-2}$. This
can be done by using the recursion relation satisfied by the
determinants $d_n$, see Eq.~(\ref{recursion-an}) with $b = g$.  We
obtain finally
\begin{eqnarray}
c_n &=& - (1 + g^2 - J_0^2 J_L^2 - 2 g \cos k) d_{n-1} 
  \label{rec-cn-eta1-eta2} \\
&&     + \left[ (1 + g^2) (1 - J_0^2 - J_L^2) + J_0^2 J_L^2 
      -2 g (1 - J_L^2) (1 - J_L^2) \cos k\right] d_{n-2}.
\nonumber 
\end{eqnarray}
Note that the recursion relation (\ref{rec-cn-eta1-eta2})
is symmetric with respect to the
exchange of $J_0$ and $J_L$.  The eigenvalues of $C_n$ are obtained by
requiring $c_n = 0$.  Setting $J_0 = J_L = \zeta$ and $n = L+1$, we
obtain Eq.~(\ref{secular-equation}).  Note that we have simplified
$\sin k$ in the denominators, hence Eq.~(\ref{secular-equation}) has a
spurious solution for $k = 0$.  A true solution with $k=0$ occurs only
when the coefficients of $d_{n-1}$ and $d_{n-2}$ in
Eq.~(\ref{rec-cn-eta1-eta2}) both vanish for $k=0$. This only occurs
when $J_0^2 = J_L^2 = 1 - g$.

\section{Perturbative analysis in the small-$g$ limit for opposite boundary
conditions} \label{App.B}

The low-energy behavior of the model with Hamiltonian (\ref{Isc}) 
can be understood analytically
in the limit $g\to 0$. For $g=0$, it is trivial to compute the 
spectrum of Hamiltonian (\ref{Isc}) in the presence of the magnetic 
boundary term (\ref{hb}), since $H$ is diagonal in the basis in which 
$\sigma_i^{(1)}$ is diagonal. Assuming $\zeta_1 = - \zeta_L = -\zeta$ and 
$\zeta > 0$, there are two family of states that control
the low-energy spectrum. First, we should consider the magnetized states 
\begin{eqnarray}
  |+\rangle &=& |1,1,1,\ldots, 1\rangle, \nonumber \\
  |-\rangle &=& |-1,-1,-1,\ldots, -1\rangle;
\end{eqnarray}
then, we should also consider the $(L-1)$ states 
(we call them kink states \cite{Sachdev-book}) 
\begin{eqnarray}
  |1\rangle_k &=& |-1,1,1,\ldots, 1\rangle, \nonumber \\
  |2\rangle_k &=& |-1,-1,1,\ldots, 1\rangle, \nonumber \\
   \ldots \nonumber \\
  |L-1\rangle_k &=& |-1,-1,\ldots,-1,1\rangle .
\label{kink-states}
\end{eqnarray}
If $H_0$ is the Hamiltonian for $g = 0$ we have ($J=1$)
\begin{eqnarray}
H_0 |\pm\rangle = - (L-1) |\pm\rangle, \qquad
H_0 |i\rangle_k = - (L-3+2\zeta) |i\rangle_k.
\end{eqnarray}
The value of $\zeta$ determines which of the states is the ground state of the 
system. For $\zeta < 1$, the ground state is doubly degenerate and spanned by
$|+\rangle$ and $|-\rangle$, while for $\zeta > 1$ the $(L-1)$ kink states 
are the lowest-energy ones. For $\zeta=1$, the magnetized and the kink states
are degenerate. We wish now to determine how this picture changes when
the perturbation 
\begin{equation}
   H_g = - g \sum_i \sigma^{(3)}_i
\label{Hg-def}
\end{equation}
is added.

\subsection{Low-field behavior}

Let us first understand the behavior for $\zeta < 1$. For $g=0$ the ground 
state is doubly degenerate. Such a degeneracy is lifted by 
perturbation (\ref{Hg-def}). 
Note, however, that $\langle \pm |H_g^n| \mp \rangle$ 
vanishes for any $n < L$. Therefore, the energy gap is proportional to 
$g^L$, in agreement with the exact results obtained in Sec.~\ref{sec5}.

Let us now consider the corrections to the ground state which are
proportional to $g$.  At order $g$, $|+\rangle$ mixes with the states
generated by $H_g|+\rangle$, so that we can write
\begin{eqnarray}
\psi_+ &=& \alpha |+\rangle + \beta |-1,1,1,\ldots,1\rangle_1 + 
    \delta |1,1,\ldots,1,-1\rangle_L +
\nonumber \\
    && + \gamma \sum_{i=2}^{L-1} |1,1,\ldots,-1,\ldots,1,1\rangle_i,
\label{correction-SG-smallg}
\end{eqnarray}
where $\beta$, $\gamma$, $\delta$ are of order $g$ and state $|\ldots
\rangle_i$ is defined so that $\sigma_j |\ldots \rangle_i = |\ldots
\rangle_i$ for $i\not=j$ and $\sigma_i |\ldots \rangle_i = -|\ldots
\rangle_i$.  Requiring $\psi_+$ to be normalized, we immediately
obtain $\alpha = 1 + O(g^2)$.  The coefficients $\beta$, $\gamma$, and
$\delta$ are fixed by the eigenvalue condition
\begin{equation}
 H_g |+\rangle = (E_0 - H_0) (\psi_+ - |+\rangle).
\end{equation}
We obtain 
\begin{eqnarray}
   \beta = {g\over 2(1- \zeta)}, \qquad
   \gamma= {g\over 4}, \qquad
   \delta= {g\over 2(1+ \zeta)}.
\end{eqnarray}
The analysis for $|-\rangle$ is analogous and leads to 
$\psi_- = T P_z \psi_+$. Once the degeneracy is lifted, the ground state 
should be an eigenstate of $T P_z$, hence it should be of the form 
\begin{equation}
 {1\over \sqrt{2} } (\psi_+ \pm \psi_-).
\end{equation}
We are now in the position to compute correlation functions on the ground
state. For the local magnetization, we obtain 
\begin{eqnarray}
m_i = 0 
\qquad \hbox{for } i=2,\ldots L-1 \nonumber \\
m_1 = -m_L = 
   \delta^2 - \beta^2 = - {g^2 \zeta\over (1-\zeta^2)^2}.
\end{eqnarray}
This result shows the the ground state does not show a local
magnetization, except at the boundaries. For $i\not=1,L$, the local
magnetization $m(i)$ vanishes as in the zero-field case.  The applied
field $\zeta$ is too small to destabilize the low-energy state. We can
analogously compute the correlation function $G(i,j)$:
\begin{eqnarray}
G(1,L) &=& 1 - 2 \beta^2 - 2 \delta^2 = 
  1 - {(1 + \zeta^2) g^2\over (1-\zeta^2)^2}, \\
G(i,L) = G(1,i)  
&=& 1 - \beta^2 - \delta^2 - 2 \gamma^2 = 
1 - {(5 + 2 \zeta^2 + \zeta^4) g^2\over 8 (1 - \zeta^2)^2}, \\
G(i,j) &=& 1 - 4 \gamma^2 = 1 - g^2/4 = m_0^2 + O(g^3),
\label{correlation-function-smallg}
\end{eqnarray}
where $1 < i \not = j < L$. Except on the boundaries,
the correlation function is equal to the square of the bulk magnetization
$m_0^2$, as in the zero-field case. 

\subsection{Large-field kink phase}

Let us now consider the large-field region in which the ground state is 
$(L-1)$ degenerate and the relevant states are the kink states
(\ref{kink-states}). The degeneracy is lifted when $H_g$ is included. 
The full Hamiltonian restricted to the subspace spanned by the kink states
has the form (we write it for $L = 5$) 
\begin{equation}
H = \begin{pmatrix} 
    E_0 & -g & 0 & 0 \\
    -g  & E_0 & -g & 0 \\
    0   & -g  & E_0 & -g \\
    0   &  0   & -g  & E_0 
   \end{pmatrix} \; ,
\end{equation}
where $E_0 = - (L-3) - 2 \zeta$.
The eigenvalues and eigenvectors of this matrix have been computed in
\ref{App.A1}. There are $L-1$ energy states 
\begin{equation}
E_m = E_0 + 2 g\cos {\pi m\over L}, 
\end{equation}
with $m = 1,\ldots L-1$. The ground state corresponds to $m = L-1$.
The corresponding eigenfunctions are 
\begin{equation}
|m\rangle = \sqrt{2\over L} \sum_{j=1}^{L-1} (-1)^j
     \sin k_m j |j\rangle_k,
\label{mstate-kink}
\end{equation}
with $k_m = \pi m/L$. Using this expression, we can compute the average of 
$\sigma_i^{(1)}$. For the average on state $m$, we obtain 
\begin{eqnarray}
&& \langle m |\sigma_1^{(1)}|m\rangle = - 
\langle m |\sigma_L^{(1)}|m\rangle = -1, \nonumber \\
&& \langle m |\sigma_i^{(1)}|m\rangle = 
     {2\over L} \sum_{j<i} \sin^2 k_m j -
     {2\over L} \sum_{j\ge i} \sin^2 k_m j,
\label{magnetization}
\end{eqnarray}
where $2\le j \le L-1$. If we now consider the ground state, 
in the large-$L$ limit, we can rewrite Eq.~(\ref{magnetization}) as 
\begin{eqnarray}
&&  m_j = 
   - 1 + {2 j\over L} - {1\over \pi} \sin {2 \pi j\over L}.
\end{eqnarray}
If we define $x = i - L/2$ and $\ell = L/2$ we obtain the more symmetric form
\begin{equation}
 m(x) = {x\over \ell} +
    {1\over \pi} \sin {\pi x\over \ell}.
\end{equation}
Analogously, we can compute the correlation function.
We obtain 
\begin{eqnarray}
&& \langle m| \sigma_1^{(1)} \sigma_L^{(1)}|m\rangle = -1, \\
&& \langle m| \sigma_1^{(1)} \sigma_j^{(1)}|m\rangle =
   - \langle m| \sigma_j^{(1)}|m\rangle, \nonumber \\
&& \langle m| \sigma_L^{(1)} \sigma_j^{(1)}|m\rangle =
   \langle m| \sigma_j^{(1)}|m\rangle, \nonumber \\
&& \langle m| \sigma_j^{(1)} \sigma_k^{(1)}|m\rangle =
   {2\over L} \sum_{i<j} \sin^2 k_m i -
   {2\over L} \sum_{j\le i<k} \sin^2 k_m i + 
   {2\over L} \sum_{i>k} \sin^2 k_m i  ,\nonumber
\end{eqnarray}
where $2\le j<k\le L-2$. Focusing again on the ground state and 
taking the limit $L\to\infty$, if $x = j - L/2$,
$y = k - L/2$, we obtain
\begin{eqnarray}
G(j,k) =
  1 - {|x-y|\over \ell} - {1\over \pi} \left|\sin {\pi x\over \ell} - 
     \sin {\pi y\over \ell}\right| 
   = 1 - |m(j) - m(k)|.
\label{g-kink}
\end{eqnarray}

\subsection{Intermediate case} \label{App.B3}

We should now discuss the behavior of the system in the intermediate
case in which kink states and ferromagnetic states are degenerate. For
$g=0$ this case corresponds to $\zeta = 1$. As discussed in
Sec.~\ref{sec5}, when the $g$ dependent term is added, the
intermediate case corresponds to $\zeta_c = \sqrt{1-g} \approx 1 -
g/2$. Therefore, to discuss the intermediate case, we set
$\zeta = 1 - g/2$ and consider the model restricted to the $(L+1)$
subspace spanned by $|0\rangle = |-\rangle$, $|i\rangle_k$
($k=1,\ldots, L-1$), $|L\rangle = |+\rangle$. If $E_0 = - (L-3) - 2
\zeta_c$ is the energy of the kink states, the Hamiltonian can be
written as (here $L=4$)
\begin{equation}
H = \begin{pmatrix}
    E_0 - g &  -g & 0 & 0 & 0 \\
     -g     & E_0 & -g & 0 & 0 \\
    0  & -g & E_0 & -g & 0  \\
    0  & 0  & -g  & E_0 & -g   \\
    0  & 0  &  0  & -g & E_0 - g 
   \end{pmatrix}\; .
\end{equation}
The eigenvalues can be computed using the results of \ref{App.A1}. 
Indeed, $H$ has the same form as matrix $\tilde{A}$ defined in 
Eq.~(\ref{matrix-Atilde-def}). Since $\delta = -g$, the secular equation
(\ref{recursion-tildean}) becomes
\begin{equation}
(1 + \cos k) {\sin k(L+1)\over \sin k} = 0,
\end{equation}
which implies
\begin{eqnarray}
k = {\pi m\over L+1} \qquad m = 1,\ldots L+1.
\end{eqnarray}
As the energy of each mode is given by $E = E_0 + 2 g \cos k$, the
ground state is obtained by taking $k=\pi$.  The ground-state energy
is $E_0 - 2 g$ and the corresponding eigenvector is simply
\begin{equation}
|GS\rangle = {1\over \sqrt{L+1}} \sum_{j=0}^L |j\rangle.
\end{equation}
It is easy to compute the local magnetization and the correlation function.
We find 
\begin{eqnarray}
m(i) &=& -1 + {2 i\over L +1} ,
\end{eqnarray}
which shows that $m(i)$ varies linearly between $1 - 2/(L+1)$ and 
$-1 + 2/(L+1)$. If we set, as usual, $x = i - L/2$ and $\ell = L/2$ we obtain
in the large-$L$ limit
\begin{eqnarray}
m(i) &=&  {x\over \ell} .
\end{eqnarray}
The two-point correlation function is also easily computed 
\begin{equation}
G(i,j) = 1 - {2 |i-j|\over L +1} = 1 - |m(i) - m(j)|.
\end{equation}

\subsection{Crossover behavior} \label{App.B4}

Let us now study the crossover behavior. We consider again the same 
basis as in \ref{App.B3}. The Hamiltonian becomes (here $L=4$)
\begin{equation}
H = \begin{pmatrix}
    E_0 - g +2 g \zeta_s/L &  -g & 0 & 0 & 0 \\
     -g     & E_0 & -g & 0 & 0 \\
    0  & -g & E_0 & -g & 0  \\
    0  & 0  & -g  & E_0 & -g   \\
    0  & 0  &  0  & g & E_0 - g  + 2 g \zeta_s/L
   \end{pmatrix} \; .
\end{equation}
The secular equation is obtained by using the results of 
\ref{App.A1}, Eq.~(\ref{recursion-tildean}).
We set $b = -g$, $d = E_0$, 
$e = E_0 - g + 2 g \zeta_s/L$, and $E = E_0 + 2 g \cos k$. 
Since $\delta = 2 \zeta_s/L - 1$, we obtain 
\begin{equation}
{4\zeta_s\over L}\left(1 - {\zeta_s\over L}\right) \sin k (L+2) + 
2 \left(1 - {2 \zeta_s\over L}\right) 
\left[1 + \left(1 - {2 \zeta_s\over L}\right) \cos k\right] 
\sin k (L+1) = 0.
\end{equation}
In the kink phase in which 
$\zeta_s > 0$, $k$ varies between $k = \pi$ 
for $\zeta_s = 0$ and $k = \pi - \pi/L$ for $\zeta_s \to \infty$. Hence, we 
write $k = \pi - z/L$. Expanding the secular equation to order $1/L^2$ 
we obtain 
\begin{equation}
4 \zeta_s z + (4 \zeta_s^2 - z^2) \tan z = 0,
\label{gap-AppB}
\end{equation}
which coincides with Eq.~(\ref{sec-etac-1}). The solutions of this equation
either solve $\tan z/2 = 2 \zeta_s/z$ or $\tan z/2 = - z/(2 \zeta_s)$. 
The ground state corresponds to the lowest value of $z$ that is a solution of 
Eq.~(\ref{gap-AppB}), hence it satisfies $\tan z/2 = 2 \zeta_s/z$.
Note the limiting values: 
\begin{eqnarray}
&& z \approx 2 \sqrt{\zeta_s} \qquad\qquad 
   \hbox{for $\zeta_s \to 0$}, \nonumber \\
&& z \approx \pi (1 - 1/\zeta_s) \qquad \hbox{for $\zeta_s \to +\infty$} .
\end{eqnarray}
We wish now to compute the ground-state eigenfunction, which we express
as $(\alpha_1,\ldots,\alpha_{L+1})$ in the basis $|j\rangle$, $j = 0,\ldots L$,
defined in \ref{App.B3}.
If $\bar{k} = \pi - k = z/L$, the coefficients satisfy the relations
\begin{eqnarray}
&& \alpha_1 (2 \cos \bar{k} - 1 + 2  \zeta_s/L) - \alpha_2 = 0, 
\label{rec-alpha-scaling}\\
&& \alpha_{j-1} - 2 \alpha_j \cos \bar{k} + \alpha_{j+1} = 0, 
\label{rec-alpha-scaling-2} \\
&& \alpha_L - \alpha_{L+1} (2 \cos \bar{k} - 1 + 2  \zeta_s/L)=0 .
\end{eqnarray}
The solution of Eq.~(\ref{rec-alpha-scaling-2}) is 
\begin{equation}
\alpha_j = e_1 \cos \bar{k} j + e_2 \sin \bar{k} j.
\end{equation}
The constants $e_1$ and $e_2$ are fixed by 
condition (\ref{rec-alpha-scaling}) and by the normalization 
condition $\sum \alpha^2_j = 1$. In the large-$L$ limit these two conditions
give
\begin{eqnarray}
e_1 &=& z \sqrt{2\over (4 \zeta^2_s + 4 \zeta_s + z^2) L}, \\
e_2 &=&  2 \zeta_s  \sqrt{2\over (4 \zeta^2_s + 4 \zeta_s + z^2) L} .
\end{eqnarray}
We can now compute $m(i)$ and $G(i,j)$.
For the local magnetization we have 
\begin{equation}
m(i) = \alpha^2_1 - \alpha^2_{L+1} + 
    \sum_{1 < j \le i} \alpha^2_j - \sum_{i < j \le L} \alpha^2_j.
\end{equation}
In the large-$L$ limit, using $\tan z/2 = 2\zeta_s/z$, we obtain 
\begin{eqnarray}
m(i) &=& {z\over z + \sin z} 
\left({1\over z} \sin {zx\over \ell} + {x\over \ell}\right),
\end{eqnarray}
where $x = i - L/2$, $\ell = L/2$. Note that the prefactor guarantees
that $m(i)$ is equal to $\pm 1$ at the two boundaries.
We can also compute the correlation
function, which satisfies the relation
\begin{equation}
G(i,j) = 1 - |m(i) - m(j)|.
\end{equation}
Correspondingly, we obtain
\begin{equation}
{\chi\over L} = {z^2 - 2 + 2 \cos z + 2 z \sin z \over 2 z (z + \sin z)},
\end{equation}
and 
\begin{equation}
{\xi^2\over L^2} = 
{z^4 + 24 + 12 (z^2 - 2) \cos z + 4 z (z^2 - 6) \sin z \over 
   48 z^2 (z^2 - 2 + 2 \cos z + 2 z \sin z)}.
\end{equation}
Let us finally consider $n_s(i)$ defined in Eq.~(\ref{defden}). For 
$g \to 0$, $n_h = 1/2$, so that 
$n_s(i) = \langle \sigma_i^{(3)}\rangle /2$. Using the definition 
we obtain
\begin{eqnarray}
n_s(i) = \alpha_i \alpha_{i+1}.
\end{eqnarray}
Substituting the explicit expression of $\alpha_i$ and 
using $\tan z/2 = 2 \zeta_s/z$, we find
\begin{equation}
Ln_s(i) =  
  {z\over (z + \sin z)}
  \left(1 + \cos{z x\over \ell} \right) .
\end{equation}
The above-reported results apply to the kink phase $\zeta_s \ge 0$. 
For $\zeta_s < 0$ the ground state is a localized state with $k = \pi + i u/L$,
where $u$ satisfies the equation
\begin{equation}
4 \zeta_s u + (4 \zeta_s^2 + u^2) \tanh u = 0.
\end{equation}
The solutions of this equation satisfy either $\tanh u/2 = - 2\zeta_s/u$ 
or $\tanh u/2 = - u/(2\zeta_s)$. Since $E = E_0 - 2 g \cosh u/L$, the ground
state is obtained by considering the largest positive solution of the
equation. It is then easy to prove that it satisfies 
$\tanh u/2 = - 2\zeta_s/u$. The solution for $\zeta_s < 0$ is then obtained
by analytic continuation of the results obtained for $\zeta_s > 0$.
It is enough to replace $z$ with $-i u$. We therefore obtain 
\begin{eqnarray}
m(i) &=& {u\over u + \sinh u}
\left({x\over \ell} + {1\over u} \sinh {ux\over \ell}\right), \\
L n_s(i) &=  &
  {u\over u + \sinh u}
  \left(1 + \cosh{u x\over \ell} \right).
\end{eqnarray}
Note that $u \approx - 2 \zeta_s$ for $\zeta_s \to -\infty$, so that 
\begin{equation}
 m(i) = {\sinh 2\zeta_sx/\ell\over \sinh 2\zeta_s}, \qquad
L n_s(i) = {2\zeta_s \cosh 2\zeta_sx/\ell\over \sinh 2\zeta_s }
\end{equation}
in this limit.

\end{document}